\documentclass[toc,aps,twocolumn,showpacs,amsmath,amssymb,superscriptaddress,prc,10pt]{revtex4-2}
\usepackage{graphicx}
\usepackage{dcolumn}
\usepackage{bm}
\usepackage{enumitem}
\usepackage{babel}
\usepackage{lineno}
\DeclareMathOperator\erf{erf}

\newcommand{\rowgroup}[1]{\hspace{-1em}#1}

\newcommand{\Iff}{\textsuperscript{55}Fe}
\newcommand{\Atfo}{\textsuperscript{241}Am}
\newcommand{\Cofs}{\textsuperscript{57}Co}
\newcommand{\epg}{$ep\gamma$}
\newcommand{\ep}{$ep$}

\begin{document}
\newpage
\clearpage

\title[]{Precision measurement of radiative neutron $\beta$-decay: \\ methodology and systematic effects}

\author{J.S.~Nico}\email[Corresponding author: ]{jnico@nist.gov}\affiliation{National Institute of Standards and  Technology, Gaithersburg, MD 20899, USA}
\author{R.~Alarcon}\affiliation{Arizona State University, Tempe, AZ 85287, USA}
\author{M.J.~Bales}\affiliation{University of Michigan, Ann Arbor, MI 48104, USA}\affiliation{Physikdepartment, Technische Universit{\"a}t M{\"u}nchen, D-85748, Germany}\author{C.D.~Bass}\altaffiliation[Present Address: ]{Le Moyne College, Syracuse, NY 13214, USA}\affiliation{National Institute of Standards and  Technology, Gaithersburg, MD 20899, USA}
\author{E.J.~Beise}\affiliation{University of Maryland, College Park, MD 20742, USA}
\author{H.~Breuer}\affiliation{University of Maryland, College Park, MD 20742, USA}
\author{T.E.~Chupp}\affiliation{University of Michigan, Ann Arbor, MI 48104, USA}
\author{K.J.~Coakley}\affiliation{National Institute of Standards and  Technology, Boulder, CO 80305, USA}
\author{R.L.~Cooper}\altaffiliation[Present Address: ]{Naval Surface Warfare Center - Crane Division, Crane, IN 47522, USA}\affiliation{Indiana University, Bloomington, IN 47408, USA}
\author{M.S.~Dewey}\affiliation{National Institute of Standards and  Technology, Gaithersburg, MD 20899, USA}
\author{S.~Gardner}\affiliation{University of Kentucky, Lexington, KY 40506 USA}
\author{T.R.~Gentile}\affiliation{National Institute of Standards and  Technology, Gaithersburg, MD 20899, USA}
\author{T.E.~Haugen}\affiliation{National Institute of Standards and  Technology, Gaithersburg, MD 20899, USA}
\author{D.~He}\affiliation{University of Kentucky, Lexington, KY 40506 USA}
\author{S.F.~Hoogerheide}\affiliation{National Institute of Standards and  Technology, Gaithersburg, MD 20899, USA}
\author{H.P.~Mumm}\affiliation{National Institute of Standards and  Technology, Gaithersburg, MD 20899, USA}
\author{B.~O'Neill}\affiliation{Arizona State University, Tempe, AZ 85287, USA}
\author{J.W.~Paster}\affiliation{National Institute of Standards and  Technology, Gaithersburg, MD 20899, USA}
\author{T.~Rao}\affiliation{National Institute of Standards and  Technology, Gaithersburg, MD 20899, USA}
\author{A.K.~Thompson}\affiliation{National Institute of Standards and  Technology, Gaithersburg, MD 20899, USA}
\author{F.E.~Wietfeldt}\affiliation{Tulane University, New Orleans, LA 70118, USA}

\collaboration{RDK~II Collaboration}\noaffiliation
\date{\today}

\begin{abstract}
In the Standard Model the free neutron decays to a proton, an electron, and an antineutrino along with a continuous spectrum of photons. In 2016 the RDK~II collaboration reported on a measurement of the photon energy spectrum and branching ratio over the range of 0.4\,keV to the 782\,keV endpoint using two different detector arrays.  In the experiment, the radiative decay photons were observed in coincidence with the decay electrons and protons. In this paper, we present details of the analysis, including the determination of the systematic corrections and uncertainties and comparison of measured particle and photon energy spectra to Monte Carlo simulations.  We conclude with approaches to improving the precision of these measurements.
\end{abstract}



\maketitle{}

\section{Overview}
\label{sec:overview}

Decay of the free neutron has a long history of contributing to the investigation of electroweak interactions in the Standard Model. Its study provides increasingly precise tests of the Standard Model and yields important inputs to cosmology and other areas of physics~\cite{Dubbers2011,Baessler2014,Vos2015,Gorchtein2023,Hayen2024}.  Neutron decay determines the Cabibbo-Kobayashi-Maskawa (CKM) matrix element $V_{\rm ud}$ through measurements of the neutron lifetime~\cite{Wietfeldt2011,Navas2024,Musedinovic2025} and decay correlation coefficients~\cite{Mund2013, Mendenhall2013,Brown2018,Navas2024}. Improved precision of the results allows comparison with $V-A$ theory that is nearing that achieved by superallowed decays~\cite{Hardy2020}. As the precision of neutron experiments improves, direct measurements of its radiative decay mode become more relevant~\cite{Ivanov2019}. In addition to the proton, electron, and antineutrino, neutron decay is always accompanied by a continuous spectrum of soft photons
\begin{equation}\label{eqn:ndecay}
	n \rightarrow p + e^- + \bar{\nu_e} + \gamma.
\end{equation}
\noindent The photons originate largely from internal bremsstrahlung of the charged decay products, and insight into this fundamental process can be understood from classical electrodynamics~\cite{Jackson1999}.

Although experimental studies of radiative beta decay in nuclei have been ongoing for many decades, the history of neutron radiative decay is more recent. It was given prominence in 1996 by Gaponov and Khafizov, who made calculations of the branching ratio and energy spectrum and also proposed an experiment to measure the decay mode~\cite{Gaponov1996a,Gaponov1996b}. In 2002, the first attempt was made to observe radiative photons from neutron decay, and a limit was placed on the branching ratio~\cite{Beck2002}. The first definitive observation was made in 2006 by the RDK~I collaboration~\cite{Nico2006,Cooper2010}. A follow-up experiment, RDK~II, improved the precision of the branching ratio and also measured the radiative photon energy spectrum between 0.4\,keV and the 782\,keV photon energy endpoint~\cite{Bales2016}. This was accomplished by using two arrays of detectors that covered a larger energy range and solid angle coverage in an optimized geometry. Due to improved statistical precision, more investigation was conducted to understand systematic effects and benchmark Monte Carlo simulations.

The dominant contribution to the branching ratio of radiative neutron decay comes from internal  bremsstrahlung from the electron; recoil order terms, including vertex bremsstrahlung, contribute less than 1\,\%. For this work, numerical calculations were performed using leading order quantum electrodynamics (QED) without accounting for effects from the finite size of the nucleon~\cite{Cooper2008}. A Coulomb correction of 3\,\% was included to the radiative partial decay rate, which was not present in prior calculations~\cite{Wilkinson1982,He2013}. Other next-to-leading order effects were not included~\cite{He2013}. These values are consistent with branching ratios from other published calculations~\cite{Gaponov2000,Bernard2004,Ivanov2013} to within 1\,\%. Further theoretical work was done to study contributions to the branching ratio at the level of $\mathcal O(\alpha^2/\pi^2)$\cite{Ivanov2017}, where $\alpha$ is the fine structure constant. 

The RDK~II experiment tested the QED calculation through its determination of the shape of the photon energy spectrum and an improved measurement of the branching ratio, but it has relevance in other areas of physics. The result was used as a low-energy test
in the discrepancy between a large body of experimental data and theoretical predictions~\cite{Bailhache2024} based on Low’s soft-photon theorem~\cite{Low1954}, and there is discussion of the infrared portion of the photon spectrum being a manifestation of the dynamic Casimir effect~\cite{Lynch2022,Ievlev2024}. If one further improves the precision, radiative neutron decay can probe additional physics. For example,  measurement of the photons' circular polarization could reveal information about the Dirac structure of the weak current~\cite{Gaponov2000, Bernard2004, Cooper2008}; improved precision of the branching ratio would constitute a test of heavy baryon chiral perturbation theory calculation~\cite{Bernard2004}; and a source of time-reversal violation may arise in a triple-product correlation between the antineutrino, electron, and photon~\cite{Gardner2012, Gardner2013}. Furthermore, as precision continues to improve in measuring correlation coefficients, the effect of radiative corrections increases in importance~\cite{Ivanov2019}.

To address any of these questions, improvement in measuring neutron radiative decay must entail a better understanding of systematic effects and the Monte Carlo simulations as well as better counting statistics. Toward that end, we present the results of the RDK~II experiment in more detail than the original publication. The methodology is the same, but here there is emphasis on the evaluation of systematic effects and comparisons of the data with simulations. In Section~\ref{sec:apparatus}, we give an overview of the apparatus used for the experiment. Section~\ref{sec:analysis} presents some details of the data and analysis including Monte Carlo (MC) methods, and Section~\ref{sec:systematics} discusses in detail the systematic effects and quantifies the corrections and uncertainties. The results are given in Section~\ref{sec:conclusion} along with a discussion of approaches for improving measurements of radiative neutron decay.

\section{Experimental Apparatus}
\label{sec:apparatus}

The experiment was carried out at the NG-6 fundamental physics end-station at the Center for Neutron Research (NCNR) at the National Institute of Standards and Technology (NIST)~\cite{ng6beamline}.  The reactor-produced cold neutron beam was guided to the experiment as in RDK~I~\cite{Cooper2010} but with increased neutron collimation to decrease backgrounds and systematic uncertainty associated with decay location. Using a calibrated \textsuperscript{6}Li-foil neutron flux monitor~\cite{Nico2006,Yue2013,Yue2018} mounted downstream of the detection region, the typical neutron flux was determined to be $1.1\times 10^8$/s.
  
\begin{figure}
 \includegraphics[width=0.48\textwidth]{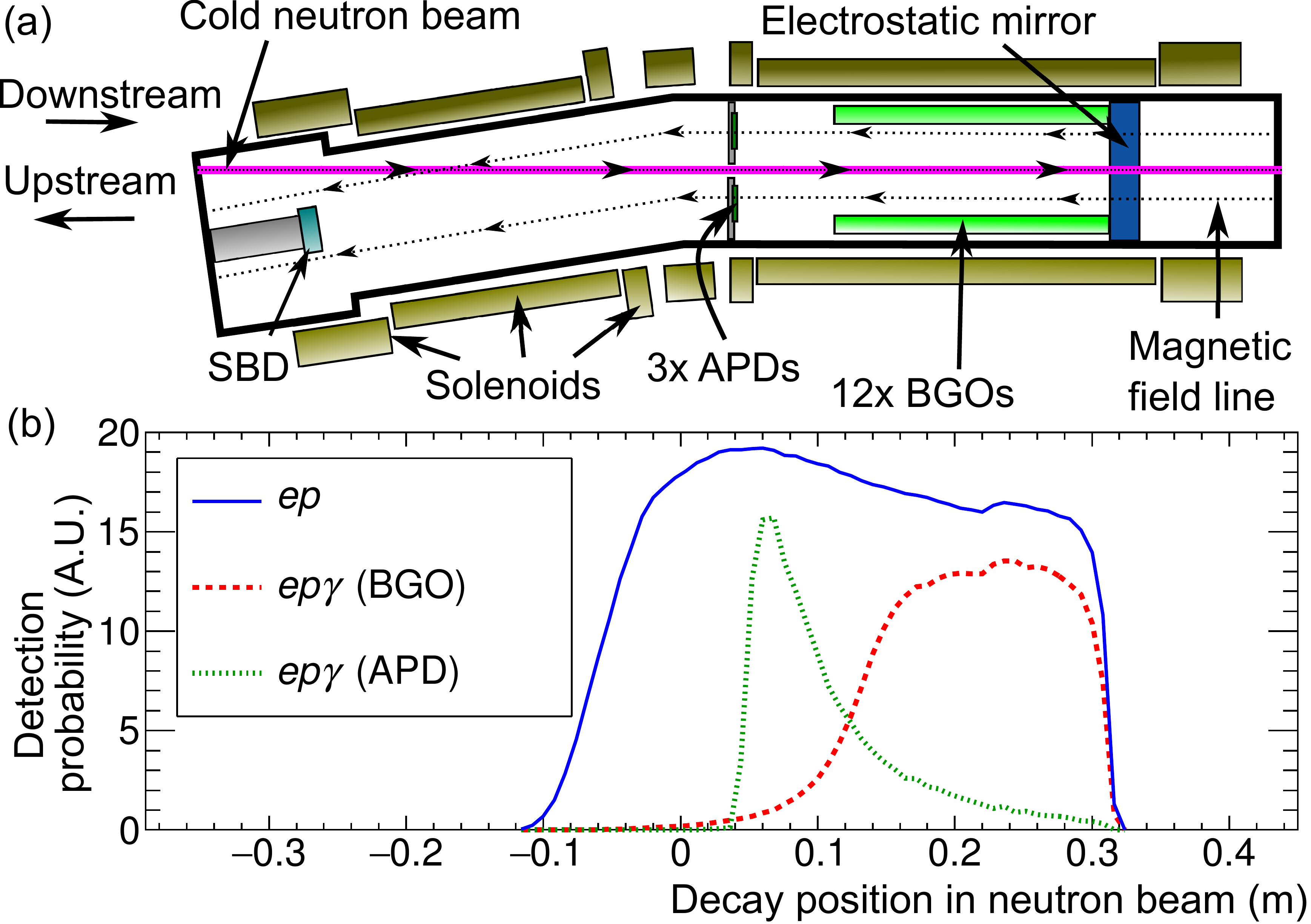}
 \caption
 {\label{fig:diagram}
(a) A cross-sectional diagram of the RDK~II detection apparatus from above. The neutron beam (pink) traveled from left to right through the active region defined by the fields (dashed lines) created by the solenoids (gold) and the electrostatic mirror (blue). Protons and electrons follow the field lines to the surface barrier detector (light blue). Radiative photons are detected by twelve bismuth germanate crystals (green) and three large area avalanche photodiodes. (b) Relative detection probability in independent arbitrary units for electron-proton (\ep) and electron-proton-photon (\epg) detection coincidence for either the bismuth germanate or direct avalance photodiode  detectors. This plot is approximately aligned with the diagram above.}
\end{figure}

The neutron beam passed through a strong magnetic field produced by a set of superconducting solenoids that were used to guide charged decay products to a detector. Several experiments that have measured neutron decay parameters have used this detection method~\cite{Byrne1990,Byrne2002,Dewey2003,Nico2006}. The detection region was defined by a 9.5\textdegree\ bend in the magnetic field and a ring of aluminum maintained at $+1400$\,V that served as an electrostatic mirror for protons (see Fig.~\ref{fig:diagram}(a)). The mirror created a $+800$\,V barrier at the center of the beam to reflect protons back to the silicon detector. The magnetic field varied from 3.3\,T to 4.6\,T over the 34\,cm distance between the bend and the mirror.

Neutrons that decayed between the mirror and the bend in the magnetic field produced electrons and protons that could be detected in the experiment.  The electrons and protons followed adiabatic helical orbits about the field lines with maximum cyclotron radii of approximately 1\,mm.  Electrons emitted in the downstream direction (see Fig.~\ref{fig:diagram}(a)) typically escaped the active detection region undetected because their kinetic energies, on the order of hundreds of keV each, were typically sufficient to overcome the electrostatic mirror.  Electrons emitted in the upstream direction followed the magnetic field to a 600\,mm\textsuperscript{2} silicon surface barrier detector (SBD)~\cite{Knoll1988,Ortec2026} in a time on the order of nanoseconds. The detectors were either 1\,mm or 1.5\,mm thick, sufficient to stop beta-decay electrons. The large thickness also reduces capacitance and hence the detector noise. Protons were detected if they were emitted in either direction because the potential on the electrostatic mirror was sufficient to reflect all of them.  The protons traveled to the SBD in a time on the order of microseconds.  The SBD was held at a $-25$\,kV potential to accelerate protons through the gold layer on its front face.  The SBD was calibrated by determining the electron endpoint energy of neutron decay from a functional fit, and its linearity was verified with radioactive source measurements.

\begin{figure}
\includegraphics[width=0.5\textwidth]{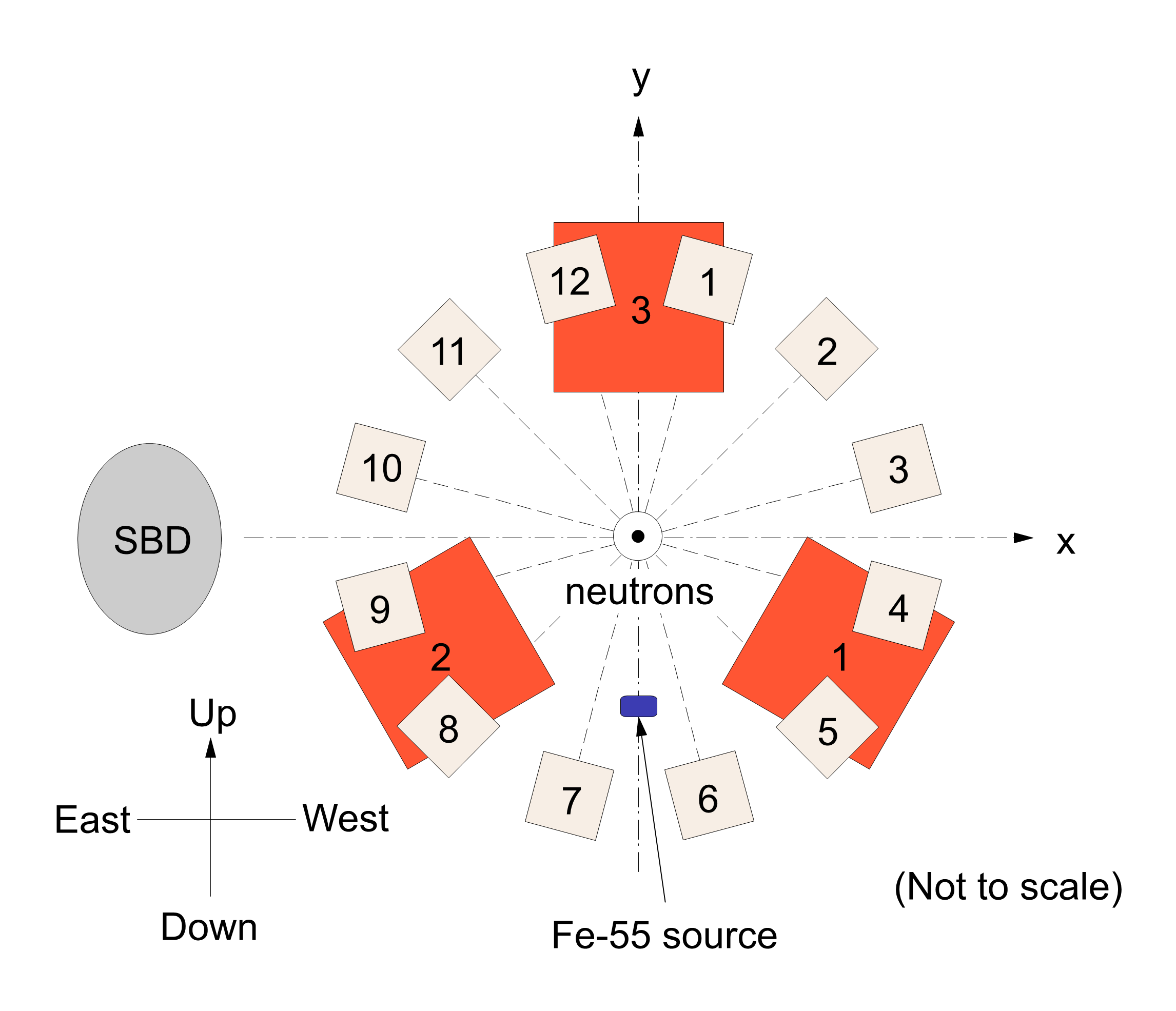}
\caption{\label{fig:Detmap} Positions of the SBD, the twelve BGO detectors, and the three APD detectors relative to one another. The view is looking upstream along the beam axis.}
\end{figure}

Two distinct photon detector arrays were employed to cover a larger energy range than in the RDK I experiment. They served to help define the fiducial decay volume around the neutron beam (see Figs.~\ref{fig:diagram} and \ref{fig:Detmap}).  The BGO detector array consisted of twelve 1.2\,cm$\times$1.2\,cm$\times$20\,cm bismuth germanium oxide (BGO) scintillator crystals optically coupled to avalanche photodiodes (APD)~\cite{Cooper2012}.  The detection range of the BGO detectors was approximately 10\,keV to 1000\,keV.  The cryogenic environment (80\,K) inside the detector served to increase the BGO scintillator light output and the APD gain while decreasing the APD noise~\cite{Gentile2007}.  There was a very low probability for false events arising from bremsstrahlung associated with charged particles striking the SBD, primarily due to the geometry of the detectors and shielding. A small correction and uncertainty for this process was determined by simulation (see Section~\ref{sys:brems}).

The bare APD array consisted of three 2.8\,cm$\times$2.8\,cm APDs that directly detected photons in the range of approximately 0.3\,keV to 14\,keV~\cite{Cooper2012}.  The APDs were oriented with their bias field parallel to the magnetic field due to previously reported issues with X-ray detection if APDs at low temperature were oriented with their bias field perpendicular to a magnetic field~\cite{Gentile2011}. 

\section{Data Analysis}
\label{sec:analysis}

\subsection{Data acquisition and reduction}
\label{subsec:DAQ}

The RDK~II experiment was carried out from December 2008 until November 2009 over eight reactor cycles, two for setup and testing and six for production data. Our method was to detect the prompt coincidence of a photon and an electron followed by a delayed proton. This requirement allows one to extract the small number of radiative photons from the large photon background at the apparatus. 

\begin{figure}[ht]
\includegraphics[width=0.48\textwidth]{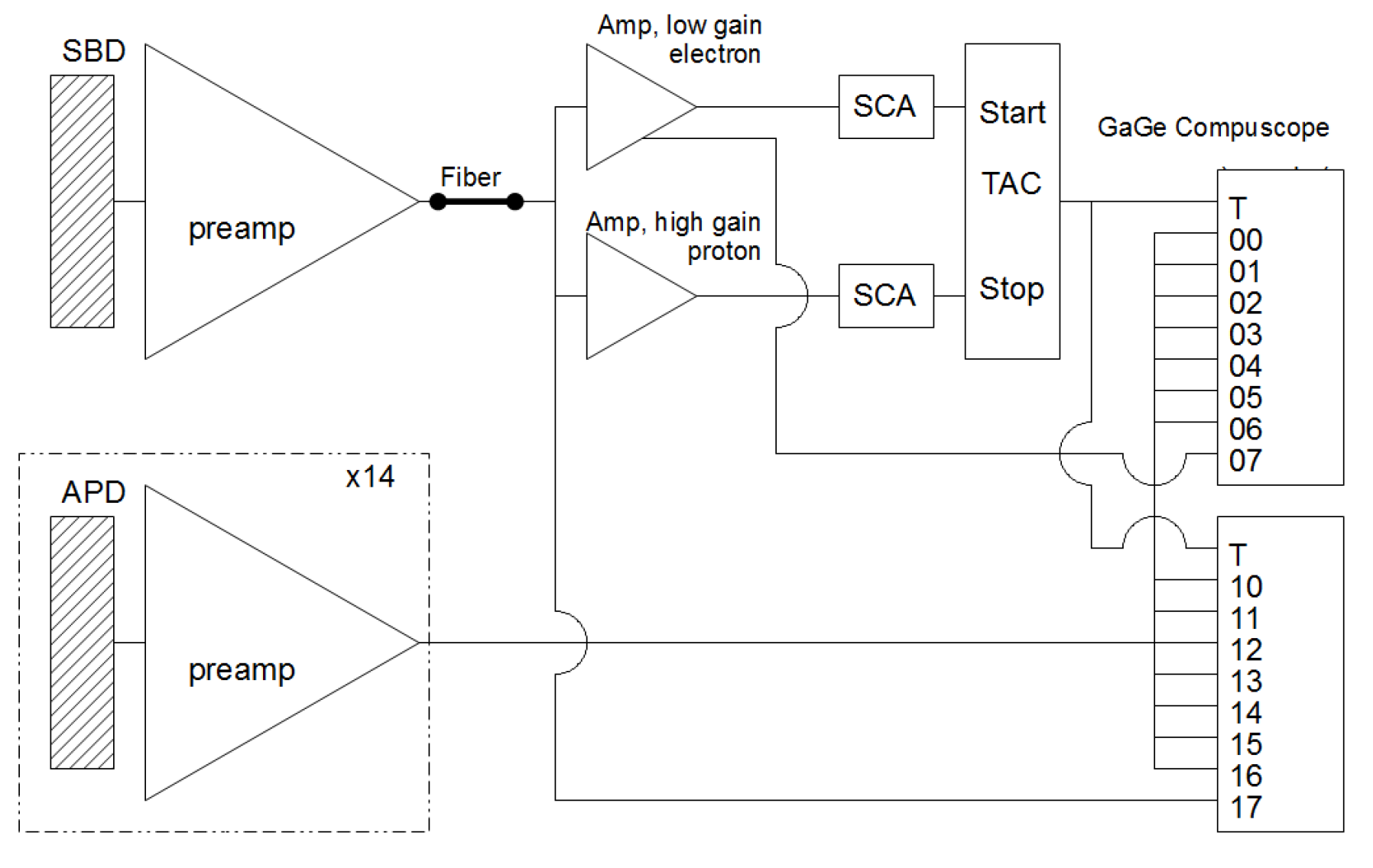}
\caption{\label{fig:electronics}
Schematic diagram of the electronics system (from Ref.~\cite{Benthesis2012}). The preamplifier for the SBD and the fiber link were fabricated specially for another experiment~\cite{Mumm2011} and employed for this work.}
\end{figure}

Data recording was initiated by two single-channel analyzers (SCA) and a time-to-amplitude converter (TAC)~\cite{Cooper2012,Benthesis2012}. The preamplifier signal from the SBD output was sent to a spectroscopy amplifier whose output was sent to two SCAs, one with an energy window for protons and the other for electrons. A signal with an energy equivalent to a $>50$\,keV electron followed by a $>7$\,keV proton (an accelerated proton), with both falling within the 25\,$\mu$s time range of the TAC, triggered data recording from the SBD and both photon detector arrays. The waveforms of all signals were recorded by two 8-channel 14-bit GaGe digitizer cards operating at 81.92\,$\mu$s full scale. The inputs were the twelve BGO signals, three APD signals, and two SBD signals (preamplifier and amplifier). Due to the limitation in the number of available channels (16), one BGO or APD was not always counted. The omitted signal was varied, so that all 15 photon channels were counted by the end of operation. The signals were digitized from 25\,$\mu$s before to 57\,$\mu$s after the electron signal with 2048 channel resolution. Each event also received a timestamp from a 25\,MHz clock. Figure~\ref{fig:electronics} shows a block diagram of the electronics.

\begin{figure}
\includegraphics[width=0.45\textwidth]{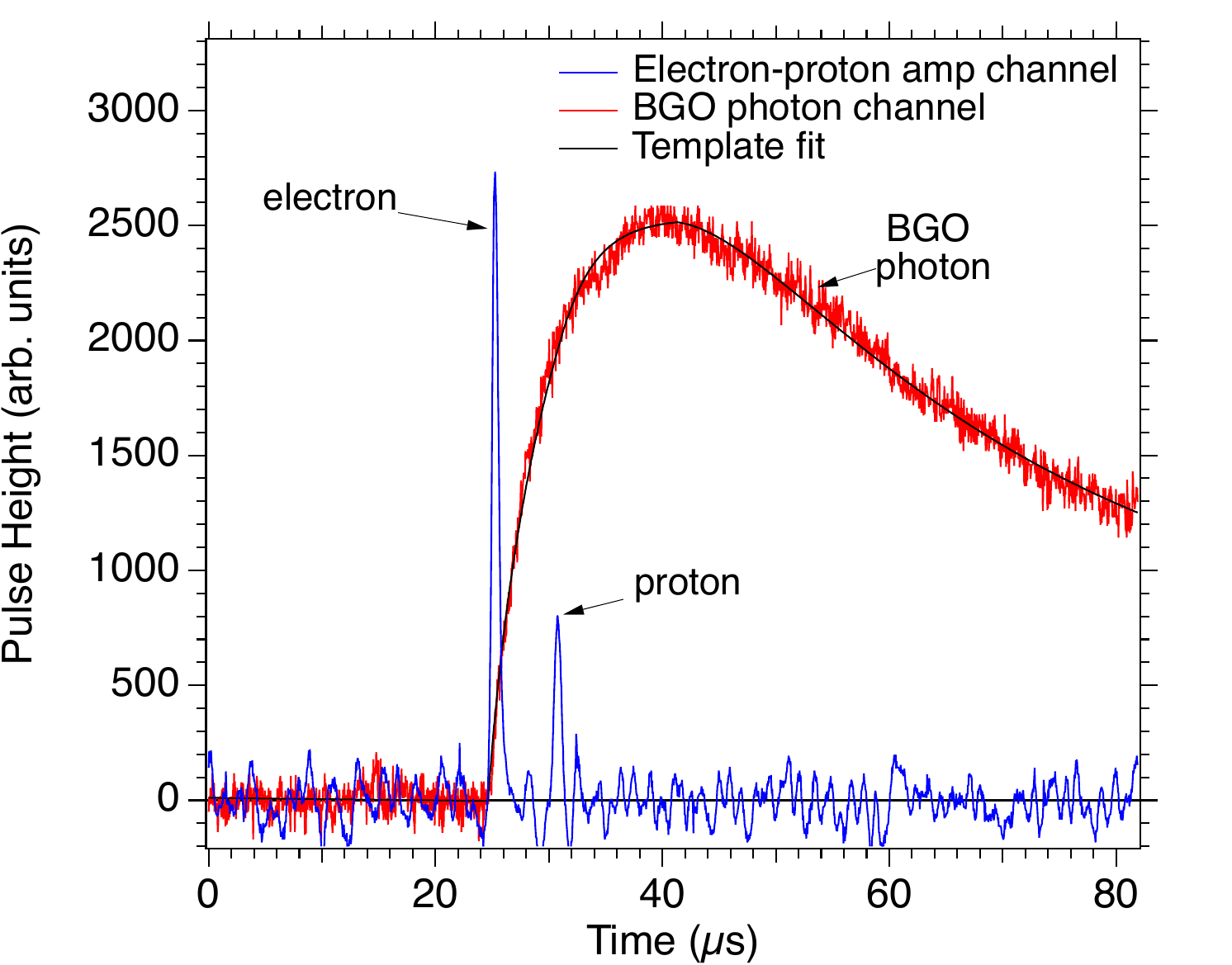}
\includegraphics[width=0.45\textwidth]{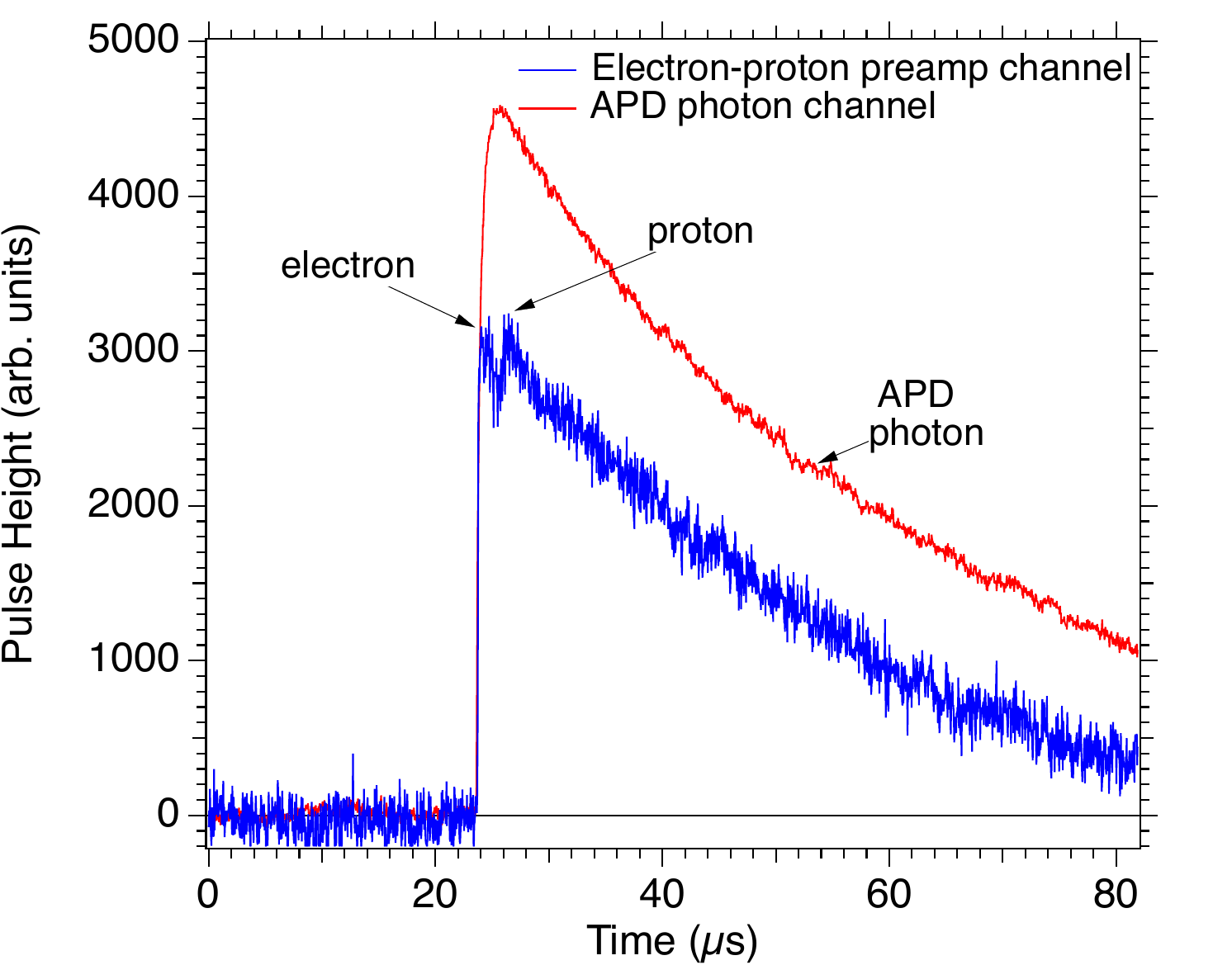}
\caption{\label{fig:waveforms} Waveforms collected for two candidate radiative decay events. (Top) The two traces are a valid electron-proton trigger and a photon event in a BGO detector occurring at approximately the same time as the electron.  The surface barrier detector has a shaped amplifier waveform for the electron and the proton (blue).  The broad preamplifier pulse from the BGO crystal shows a slow rising photon signal (red). A template fit of the BGO photon pulse is shown (black). (Bottom) The two traces are a valid electron-proton trigger and a photon event detected directly in an APD detector occurring at approximately the same time as the electron. In this case, the preamplifier signal of the surface barrier detector (blue) is shown for informational purposes; the APD analysis still used the shaped amplifier signal.  The preamplifier pulse from direct detection with an APD (red) illustrates its faster response. } 
\end{figure}

The digitized waveforms were analyzed offline to extract the energy and timing information for all the registered events. Figure~\ref{fig:waveforms} shows waveforms for a candidate radiative decay event in a BGO and an APD detector. The waveforms from the SBD and the photon detector events were fit to functional forms to extract their pulse heights.  The electron portion of the SBD waveform was fit to a Gaussian function with a linear background, and the proton portion was fit to a Gaussian function with a linear and exponential term. The exponential term was added to account for proton events that arrive quickly and whose energy may be affected by the tail of the electron waveform. The amplitude of each fit is proportional to the energy, and the position of the peak height was registered as the arrival time. Candidate photon events from BGO were fit to an empirical template function $\mathcal{F}(t)$ with five free parameters that was created using a data set of photon calibration events. The function has the form
\begin{equation}\label{eqn:template}
	\mathcal{F}(t) = (a + bt) + A\mathcal{T}[(t-t_o)s],
\end{equation}
\noindent where $\mathcal{T}(t)$ is the template, $A$ is the amplitude, $t_o$ is a timing offset, $s$ is a scaling factor, and $(a+bt)$ is an amplitude offset that allows for linear variation. The amplitude $A$ is used as a measure of a quantity proportional to the photon energy, and the timing offset $t_o$ is used for the time of the event.

Once the events were parameterized, cuts were applied to the parameters to produce the final data set; Table~\ref{tab:cuts} gives the regions for the cuts. In addition to producing the final result, the cuts were varied to study systematic effects.

\begin{table}[h]
\caption{\label{tab:cuts} Acceptance regions for the final set data. The cut on waveform identification is discussed in Section~\ref{sys:WaveformID}}
\begin{ruledtabular}
\begin{tabular}{c}
13\,keV $<$ proton energy $<$ 31\,keV                   \\
100\,keV $<$ electron energy $<$ 757\,keV               \\
2\,$\mu$s $<$ proton time-of-flight $<$ 25\,$\mu$s      \\
10\,keV $<$ BGO photon energy $<$ 780\,keV              \\
0.4\,keV $<$ APD photon energy $<$ 14\,keV               \\
-10,000 $<$ waveform identification $<$ 100               \\
\hspace{8mm}        BGO photon multiplicity $\leq 2$    \\
\hspace{8mm}        APD photon multiplicity $= 1$               \\
\hspace{8mm}        proton multiplicity $=1$               \\
\end{tabular}
\end{ruledtabular}
\end{table} 

The total live time of the data acquisition was 87.5 days, and the final data set consisted of $2.2\times 10^{7}$ electron-proton ($ep$) events for which about 20,000 and 800 coincident radiative photons were detected with the BGO and APD detectors, respectively. Data were collected in grouped ``series'', and each series consisted of runs that were typically about two hours in length. Some data runs were eliminated from the analysis, typically for their small size or the detection of an experimental problem during operation.

\subsection{Energy spectrum and branching ratio}

\begin{figure}
\includegraphics[width=0.48\textwidth]{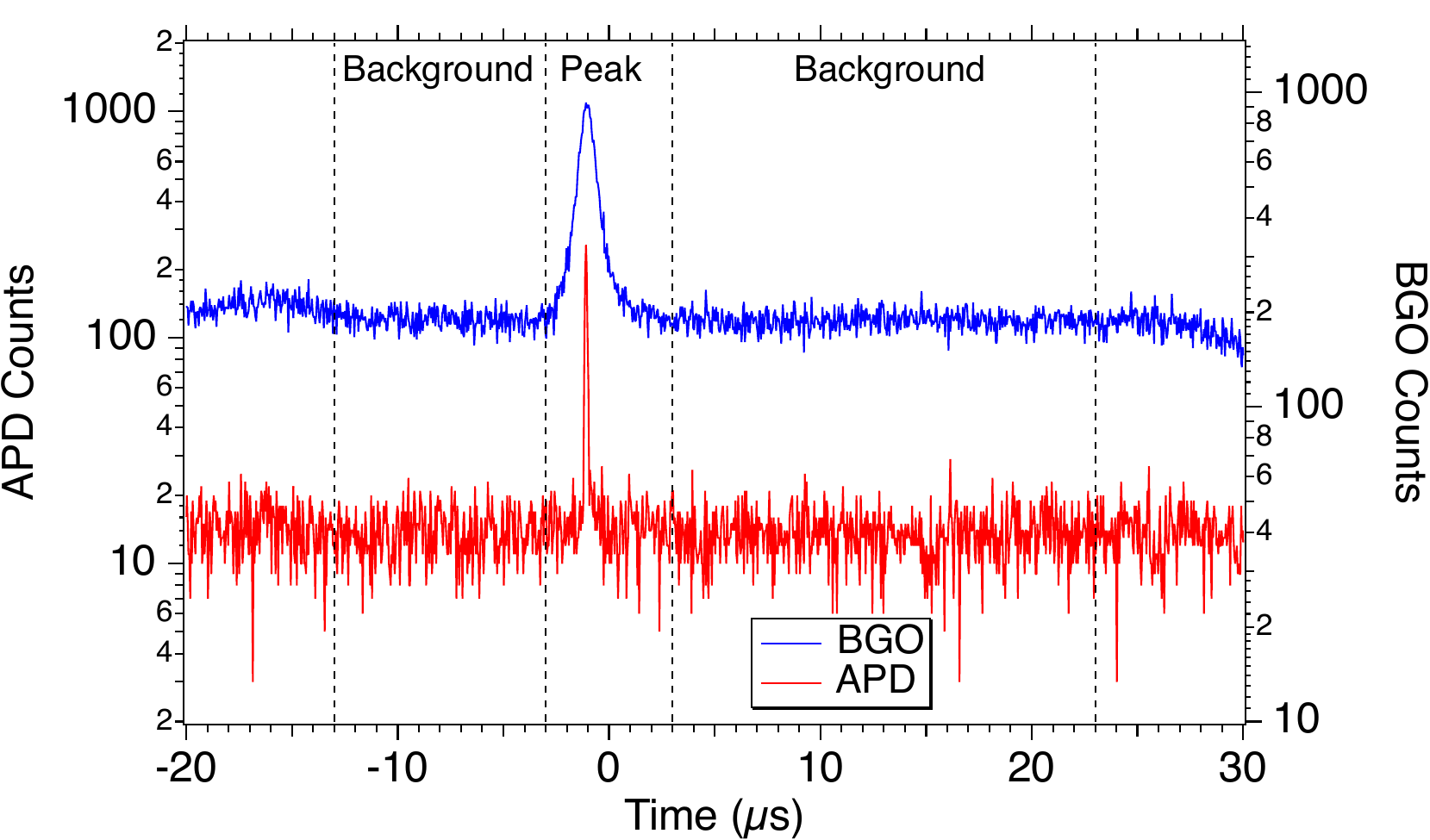}
 \caption {\label{fig:egTime} The timing spectrum for the difference between electron and photon detection in coincidence with a delayed proton. The offset from zero is due to electronic delays. The central peak arises from radiative photons, which are detected nearly simultaneously with the electrons, while the flat regions represent sources of constant, uncorrelated photon background.  The response of the APD detectors (red) was significantly faster than the BGO detectors (blue) and resulted in a narrower timing peak. The black dashed lines show the coincidence and background windows for the BGO detectors only; the comparable windows for the APDs are given in Section~\ref{sys:egtiming}.}
 \end{figure}

With the extracted timing and energy parameters for all events, cuts were made on the data to produce histograms of the difference in arrival times of the electrons and photons. For each individual photon detector, the background was determined by using the pre-peak and post-peak photon backgrounds found in the electron-photon timing spectrum (see Fig.~\ref{fig:egTime}). To obtain the radiative photon energy spectra for the BGO and APD detector, one subtracts the energy spectra of the backgrounds from the energy spectra of the coincidence peak.

To determine the branching ratio, a ratio was formed for both the experimental data and the MC simulation by dividing the detected rate of electron, proton, and photon coincidences $r_{ep\gamma}$ by the detected rate of electron and proton coincidences $r_{ep}$.  This ratio of rates $R = r_{ep\gamma} / r_{ep}$ serves two purposes: it is independent of the neutron rate and some potential systematics associated with the detection of electrons and protons cancel.  The ratio for the integrated experimental data $R_{\textrm{exp}}$  and the data from the integrated MC simulation $R_{\textrm{sim}}$ can then be compared. Because $R_{\textrm{sim}}$ depends on the theoretical branching ratio $B_{\textrm{theory}}$, an experimental branching ratio can be extracted 
\begin{equation}\label{eqn:result_eqn}
B_{\textrm{exp}}=B_{\textrm{theory}} R_{\textrm{exp}} / R_{\textrm{sim}}.
\end{equation}
\noindent Additional details and the extraction of the final results are found in Section~\ref{subsec:results}.

\subsection{Simulations and Monte Carlo methods}
\subsubsection{Simulations}

To compare the experimental data with the results from QED calculations, it was necessary to use a detailed simulation to model both the complex geometry and the nonlinear energy response of the photon detectors. The simulation requires a neutron decay generator, decay product transport, and a detailed geometric description of the relevant parts of the apparatus. Details of the MC simulations are found in Ref.~\cite{BalesThesis}, and only an outline is given here.

In the simulation, initial momenta and positions for the neutron decay products were created randomly using a leading-order QED event generator~\cite{Cooper2008} and a simulation of the neutron beam profile. Protons, electrons, and photons were transported by a Runge-Kutta algorithm in a model of the geometry of the detection region with \textsc{Geant4.9.6.p02}~\cite{geant4,BalesThesis}.  Magnetic and electric fields were interpolated from simulated field maps of the apparatus. \textsc{Geant4.9.6.p02} was also used to determine the energy deposited in the detectors, including any secondary radiation or backscattering produced.  Models of detector energy response and energy resolution were also incorporated. 

An accurate map of the static electric and magnetic fields was essential to track the helical orbits of the proton and electron from the point of decay to the SBD. The magnetic field generated by the superconducting solenoid was calculated using the
Biot-Savart package~\cite{BiotSavart,disclaimer}, which integrates the Biot-Savart equation along each coil of the magnet. The calculated field compared well with a measurement of the field and the historical use of the same magnet~\cite{Nico2005}.

Electric fields were generated for the electrostatic mirror and the SBD. The fields were computed using a finite element analysis code~\cite{Cooper2008} written in IGOR Pro~\cite{IgorPro} and using COMSOL~\cite{COMSOL}. All fields were linearly interpolated. Cubic interpolation was confirmed to have a minimal effect on the simulation result and had significantly slower performance. The possibility that the voltage bias on the APD array could have influenced the transport of charged particles was also investigated. The field was modeled in COMSOL, and the contribution was determined to have a negligible impact on the simulated results.

\subsubsection{Comparison with data}

\begin{figure*}

\includegraphics[width=0.32\textwidth]{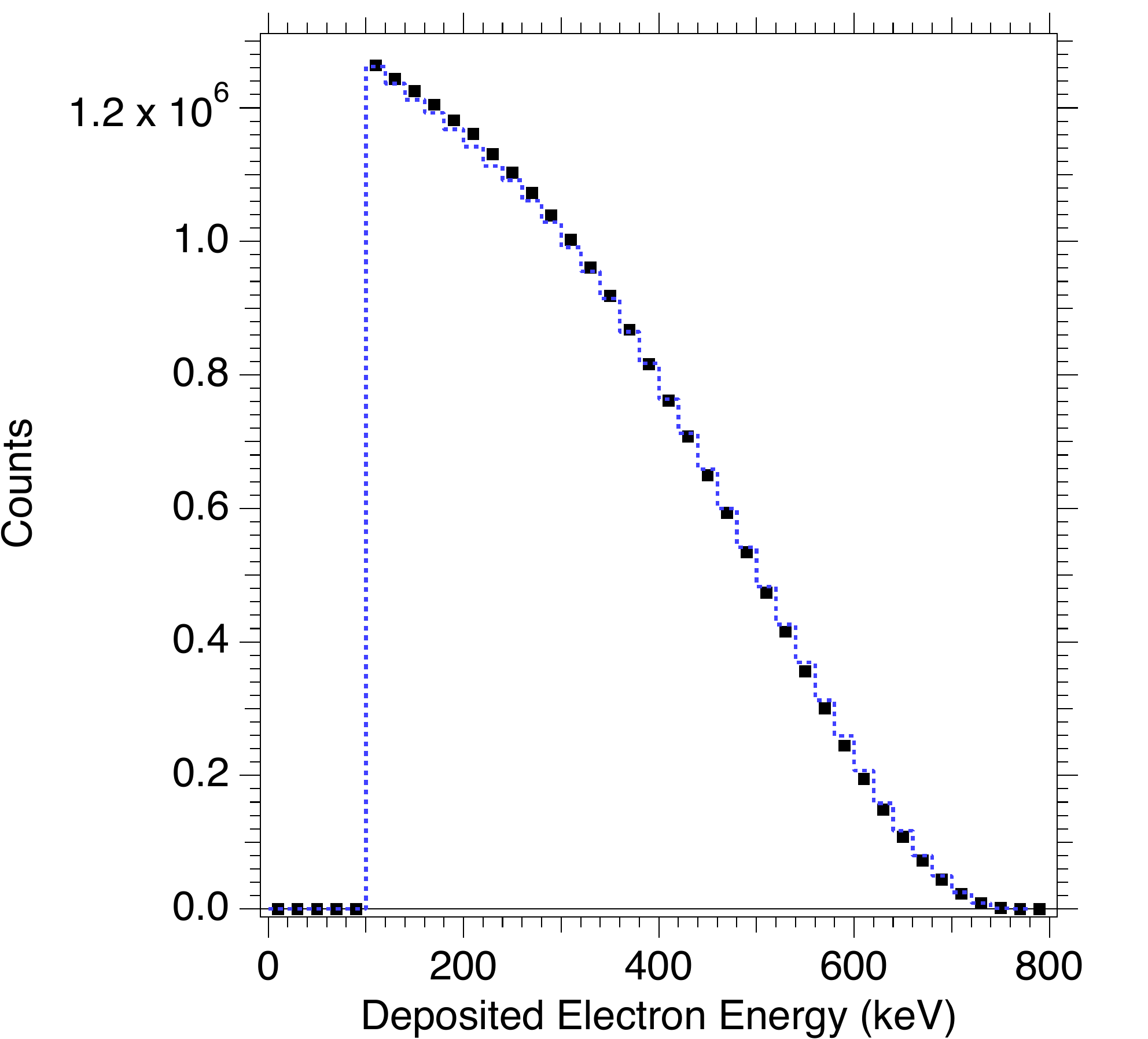}
\includegraphics[width=0.32\textwidth]{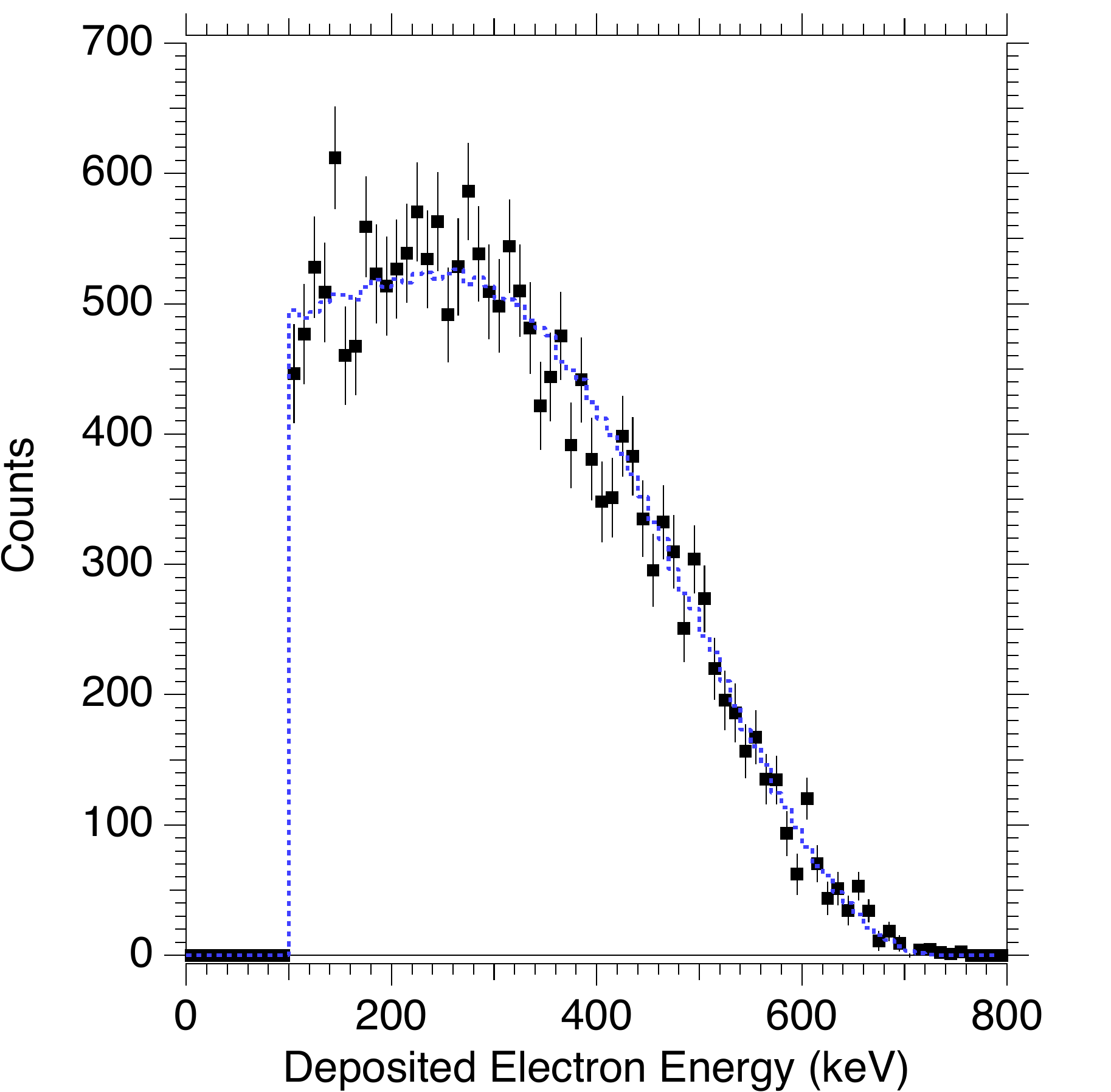}
\includegraphics[width=0.32\textwidth]{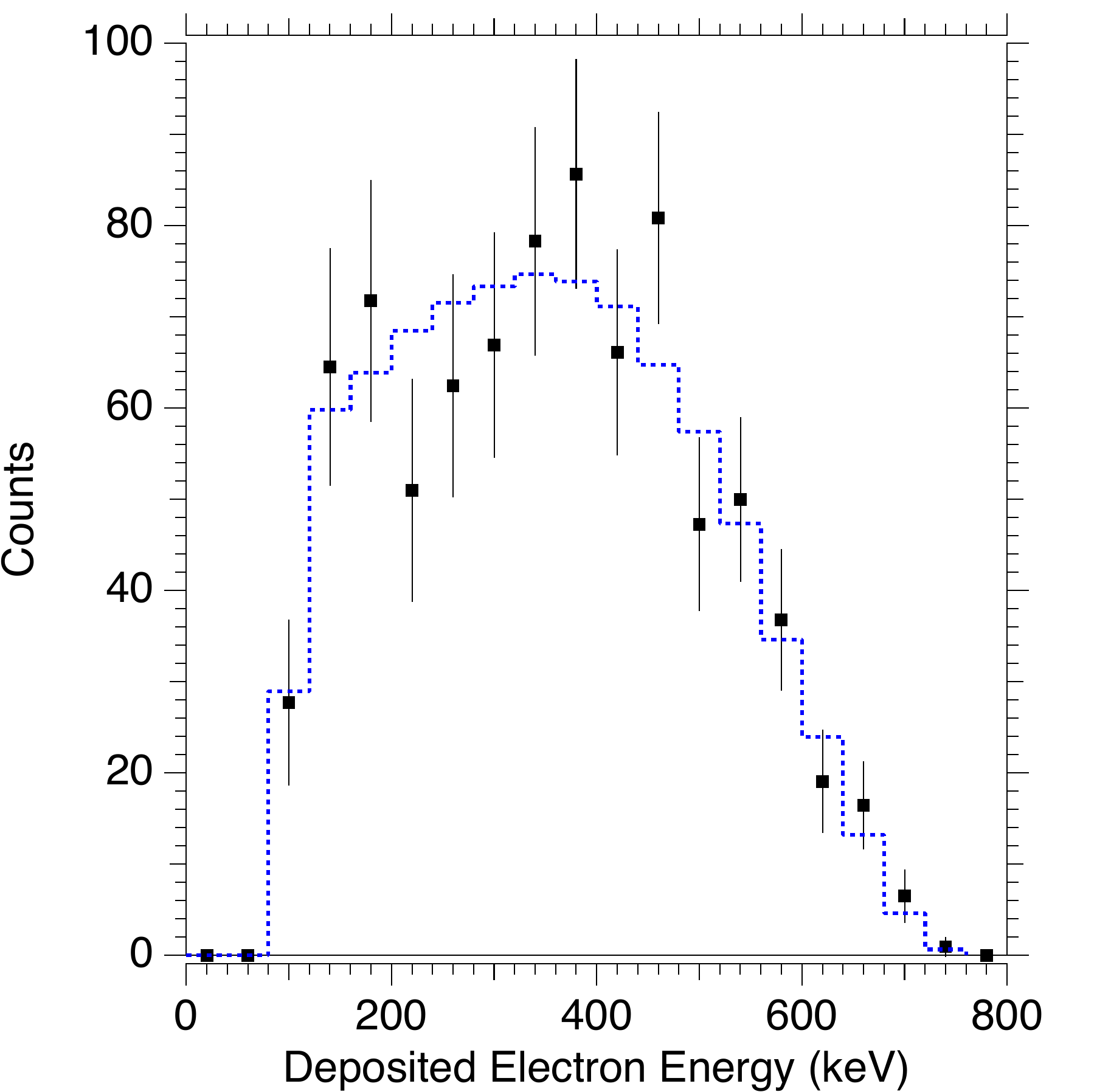}
\includegraphics[width=0.32\textwidth]{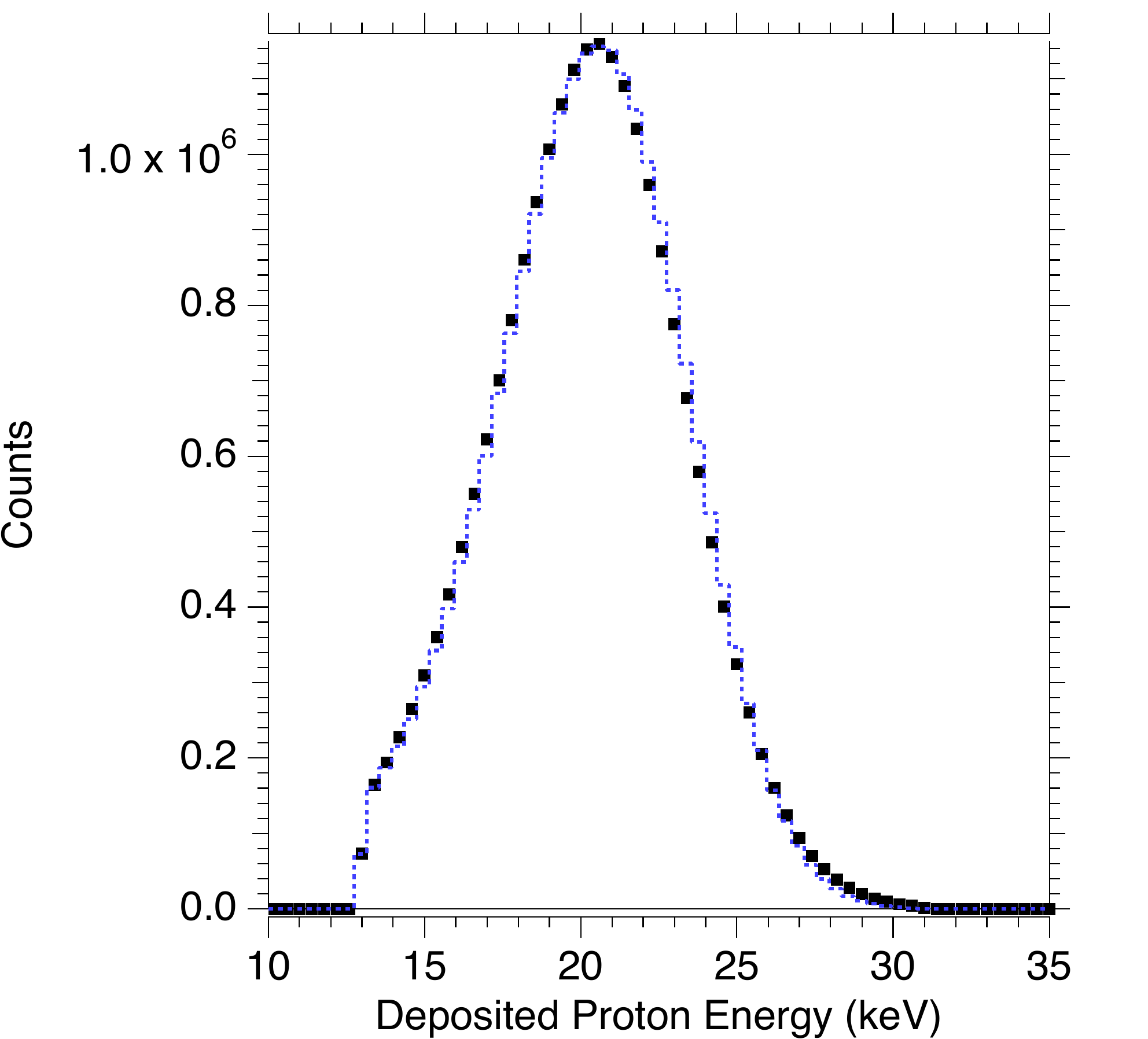}
\includegraphics[width=0.32\textwidth]{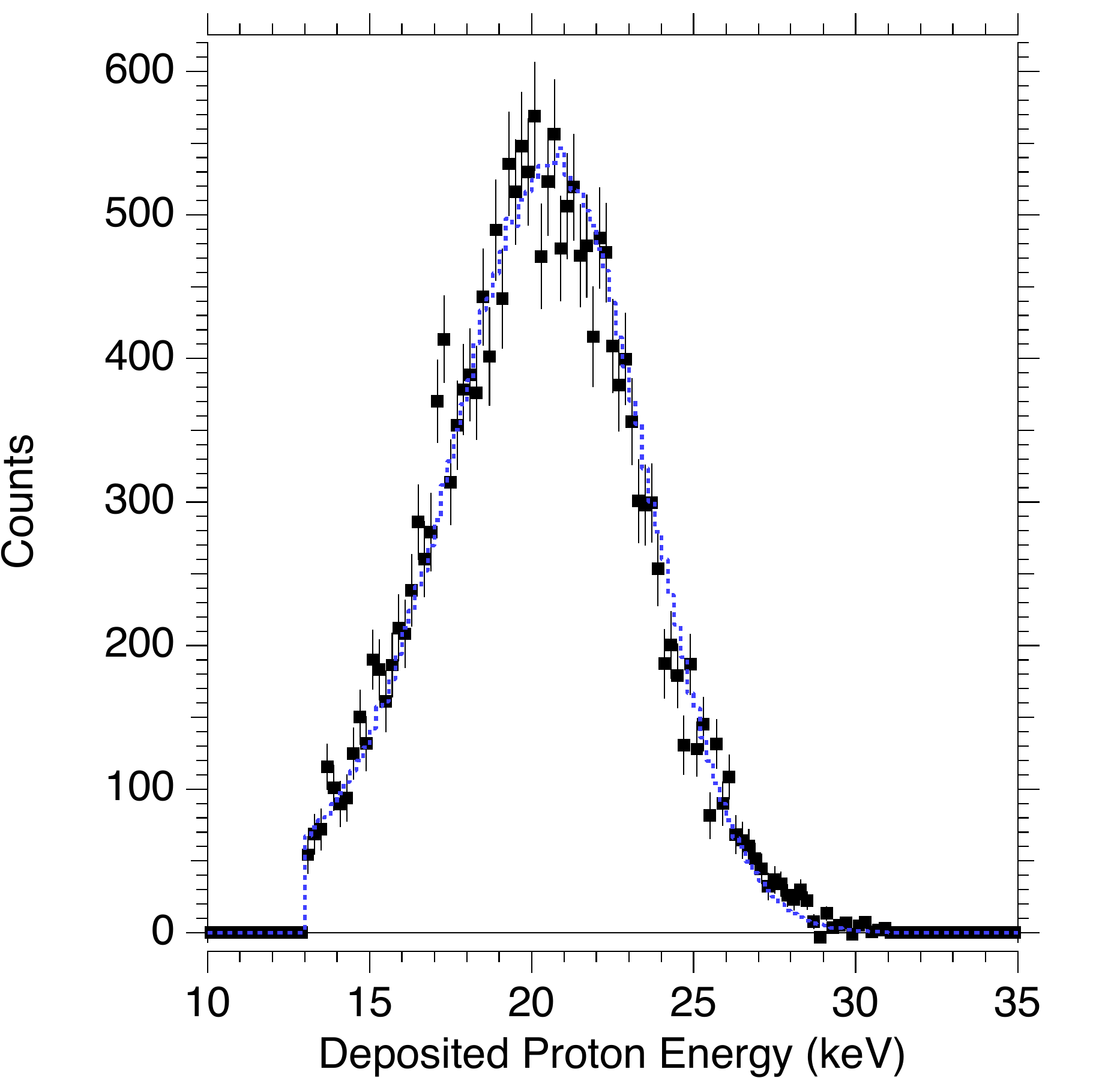}
\includegraphics[width=0.32\textwidth]{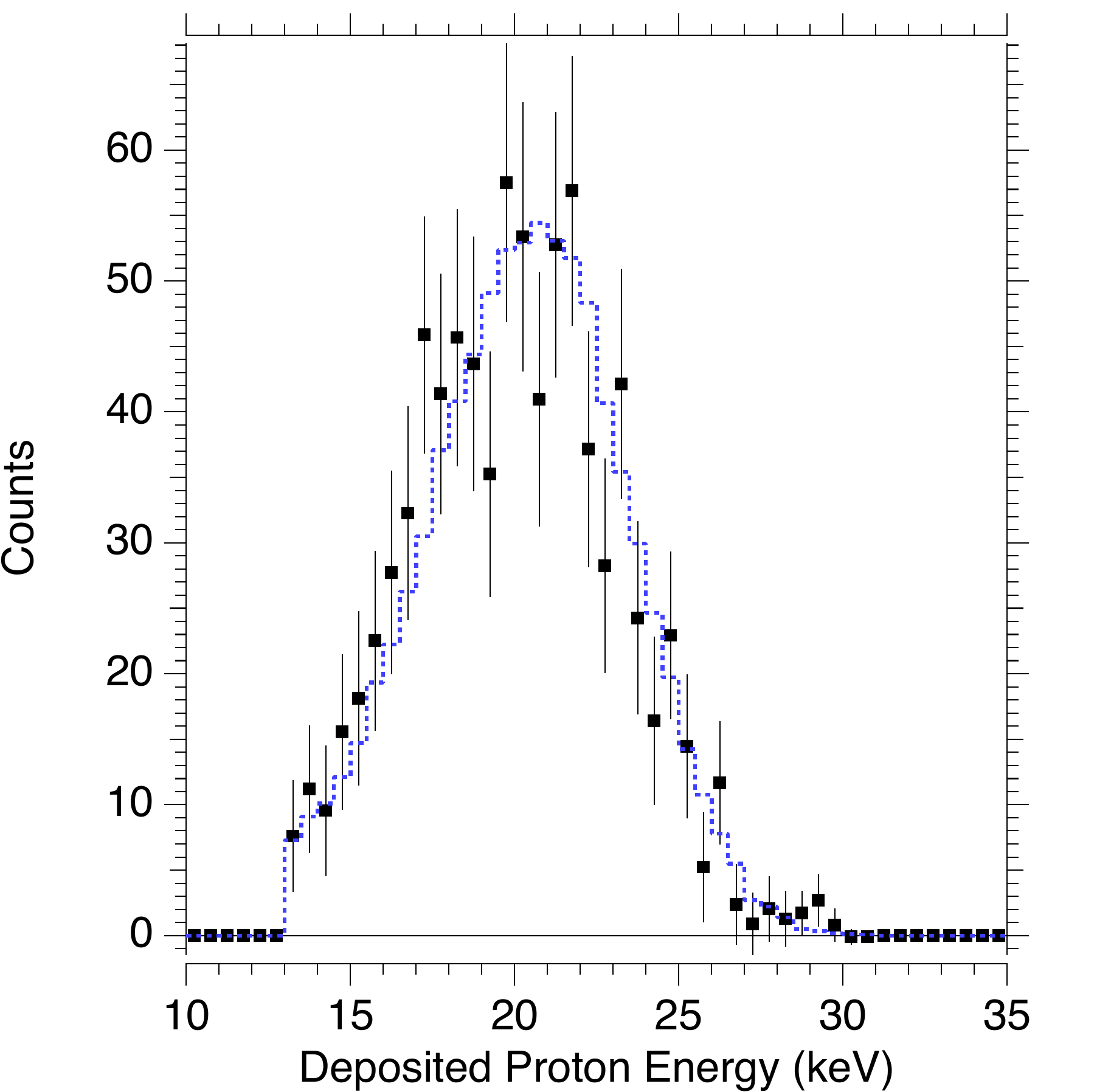}
\includegraphics[width=0.32\textwidth]{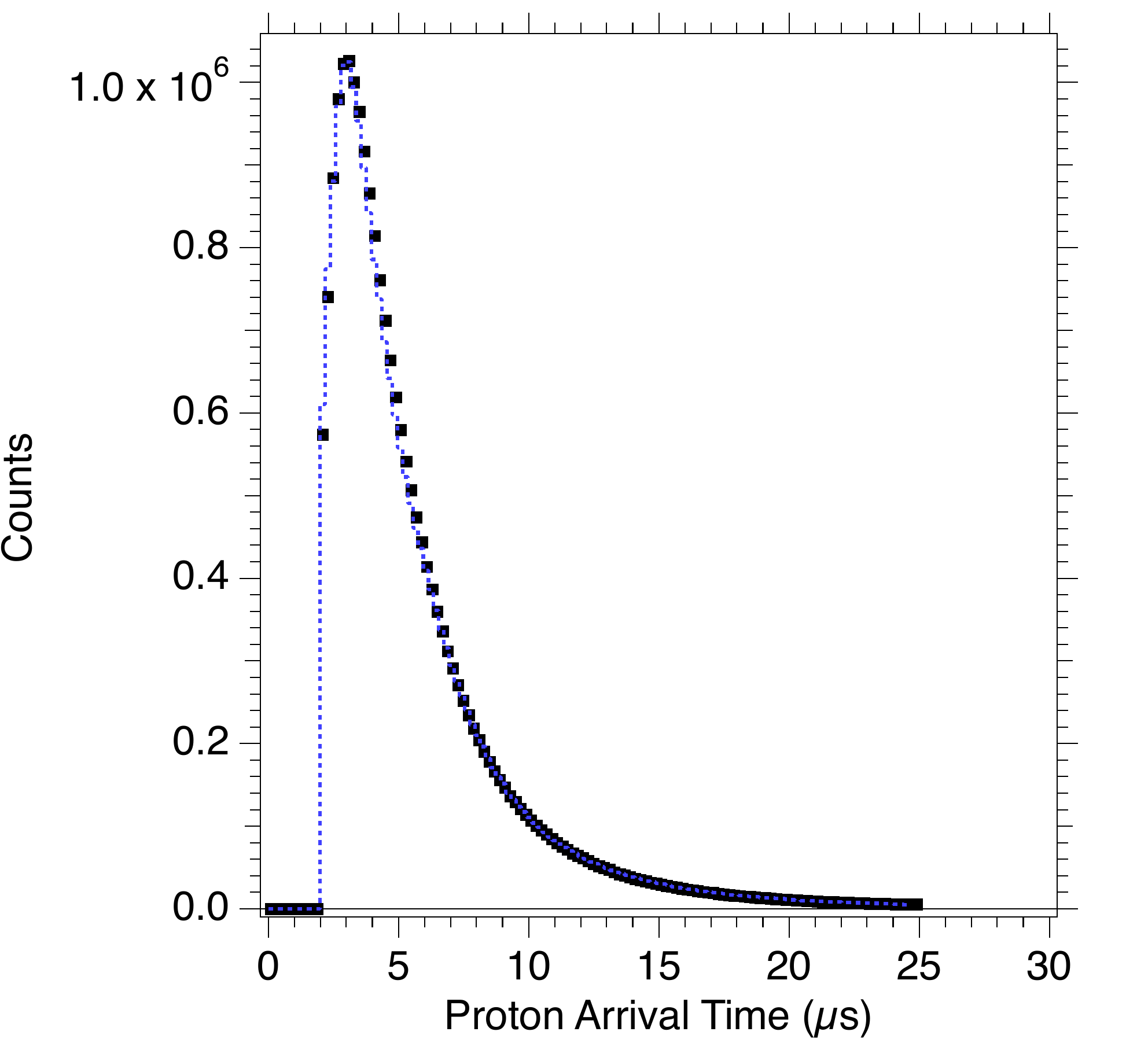}\includegraphics[width=0.32\textwidth]{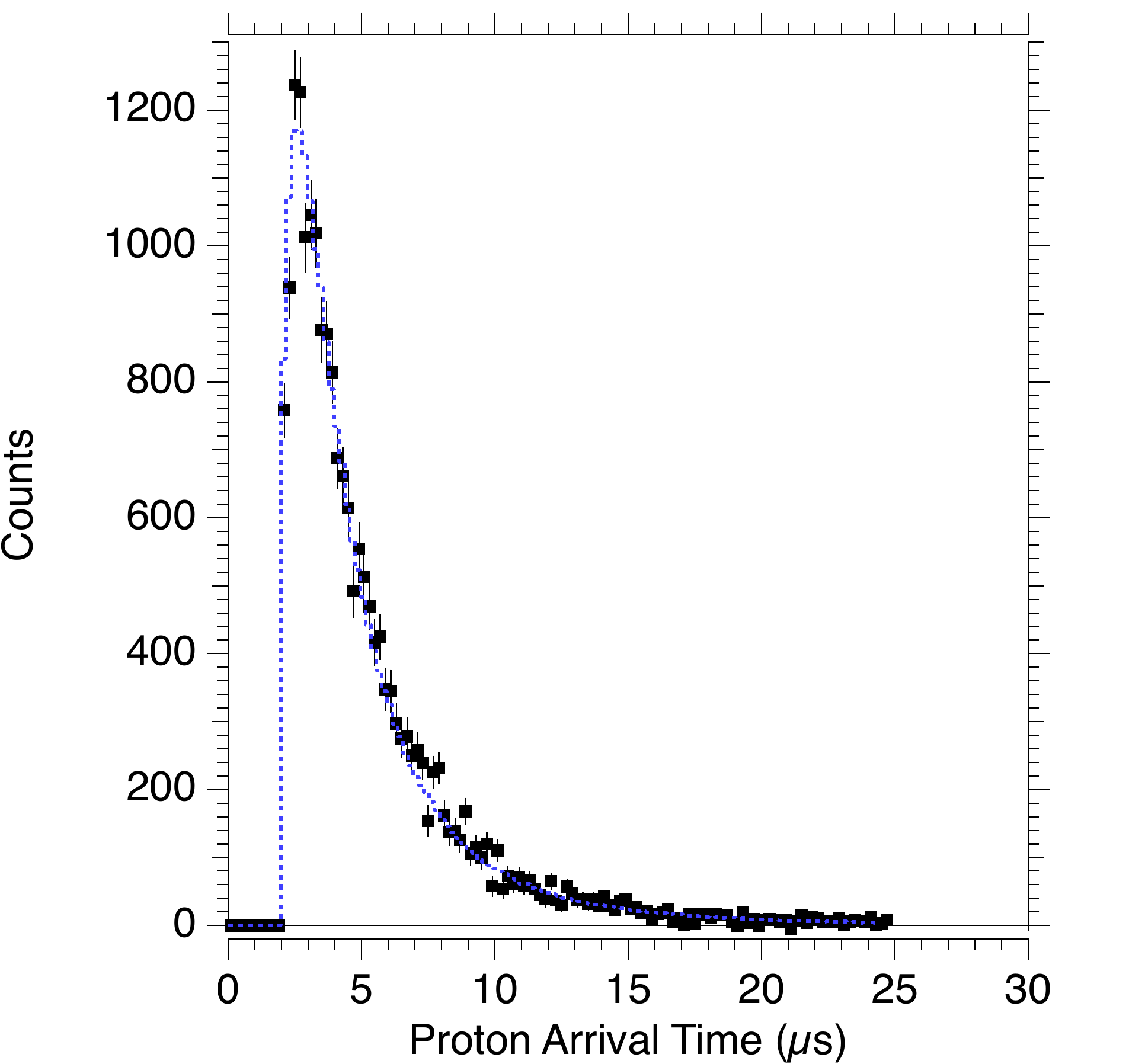}
\includegraphics[width=0.32\textwidth]{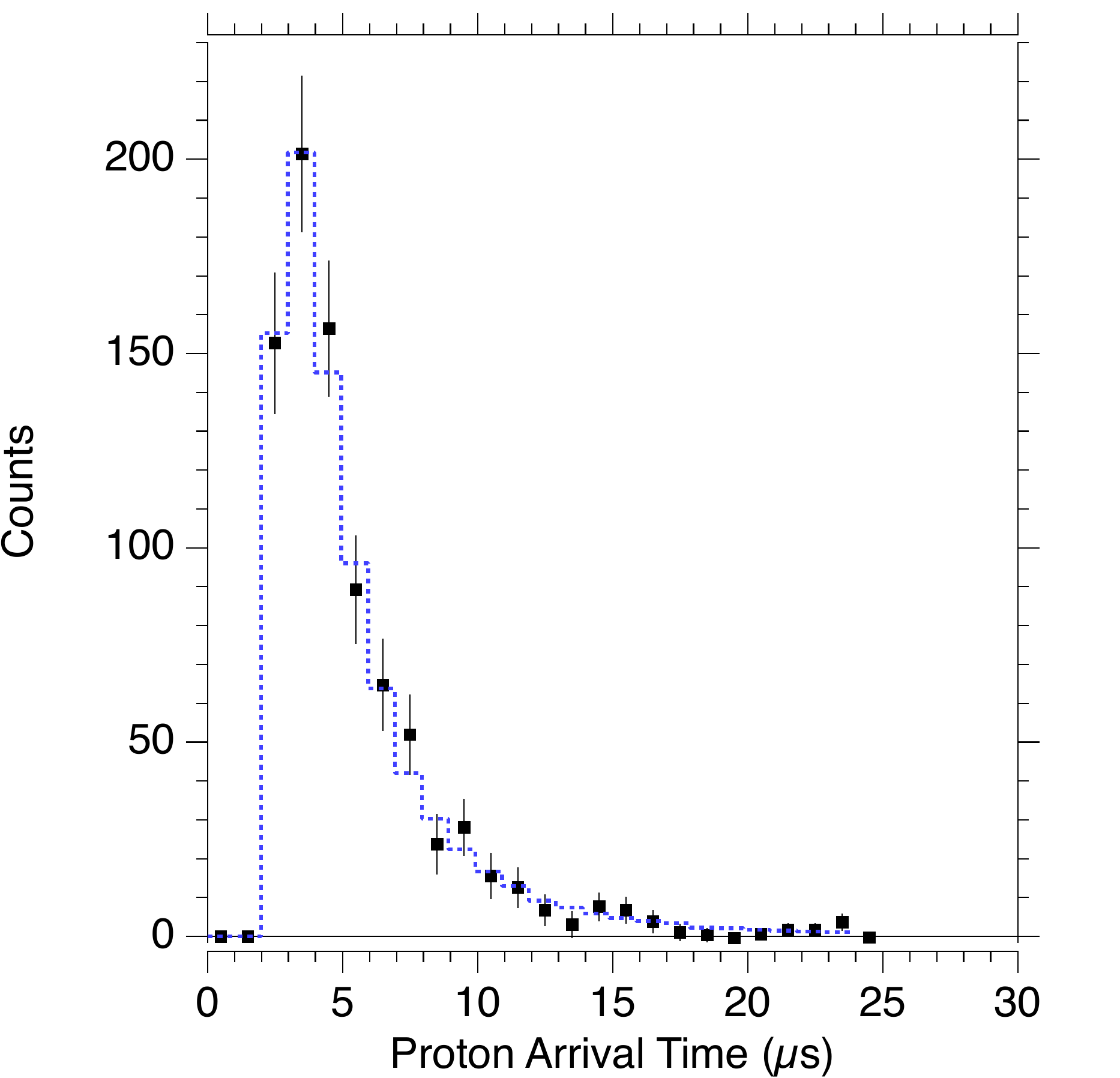}

\caption {\label{fig:MC_Data}
Comparison plots of data and the MC simulation. The top horizontal plots are the electron energy spectra; the middle horizontal plots are the proton energy spectrum; and the bottom horizontal plots are the proton arrival time. The first column of plots is events with $ep$ only; the second column of plots is $ep$ events with BGO photons; and the third column of plots is $ep$ events with APD photons. The black squares are the data and the dashed blue lines are the MC simulation. The regions that are plotted are determined by the acceptance regions of Table~\ref{tab:cuts}. The error bars represent the statistical uncertainties only.}
\end{figure*}

Benchmarking the MC against measurable quantities from the apparatus demonstrates the validity of many aspects of the MC. In addition to the photon energy (a goal of the experiment), three of the key measurable quantities are the electron energy spectrum, the proton energy spectrum, and the arrival time of a proton after an electron trigger. 
Figure~\ref{fig:MC_Data} shows those plots using the final data set after the cuts from Table~\ref{tab:cuts} were imposed. Three sets of data were compared: $ep$ events only ({\it i.e.,} no photon requirement), $ep$ events with a BGO photon, and $ep$ events with an APD photon. The simulation results were placed on the same plot using only a common scaling factor obtained from the total number of counts in each of the three data sets. The exception is the proton energy spectra, which required an additional free parameter of the width to account for the SBD detector resolution.

\begin{table*}
\footnotesize
\begin{ruledtabular}
\begin{tabular}{>{\quad}lccccc}
\footnotesize                       & BGO               &BGO                & APD               & APD               & Section \\
\footnotesize                       & Correction(\%)    & Uncertainty(\%)   & Correction(\%)    & Uncertainty(\%)   & \\
\hline
\rowgroup{Photon detectors}         &                   &                   &                   &                   & \\
Photon energy calibration           & -			        & 0.6			    & -                 & 1.3               & \ref{sys:photoncal} \\
Energy response                     & -                 & 2.6               & -                 & 10                & \ref{sys:enresponse} \\
Multiple photons                    & 0.4               & 0.2               & 0.1               & 0.1               & \ref{sys:multiplephotons} \\
\rowgroup{SBD detector}				& 			        &                   &                   &                   & \\
Electron energy calibration         & -                 & 0.2               & -                 & 0.5               & \ref{sys:SBDCal} \\
Proton energy calibration           & -                 & 0.5               & -                 & 0.4               & \ref{sys:SBDCal} \\
Waveform identification & -                 & 2.2               & -                 & 0.4               &	\ref{sys:WaveformID} \\
\rowgroup{Timing cuts}              &                   &                   &                   &                   & \\
Electron-proton timing              & -                 & 0.5               & -                 & 0.6               & \ref{sys:eptiming} \\
Electron-photon timing              & -                 & -                 & -                 & -                 & \ref{sys:egtiming} \\
\rowgroup{\epg\ backgrounds}   &                   &                   &                   &                   & \\
Electron bremsstrahlung             & -0.8              & 0.1               & -                 & -                 & \ref{sys:brems} \\
Non-decay background                & -1.0              & 1.0               & -0.4              & 0.4               & \ref{sys:nondkbkgd} \\
\rowgroup{Simulation}               &                   &                   &                   &                   & \\
Model registration                  & -                 & 2.8               & -                 & 3.4               & \ref{sys:model} \\
Statistics                          & -                 & 0.1               & -                 & 0.4               & \\
\hline
\rowgroup{Total Systematic}         & -1.4              & 4.6               & -0.3              & 11                & \\
\end{tabular}
\end{ruledtabular}

\caption{\label{table:Systematics}
Summary of the systematic corrections and relative standard uncertainties in the measured branching ratio. Each systematic effect is considered to be independent, and the total systematic uncertainty is the quadrature sum of the individual values. Note that ``-" indicates that the systematic corrections are less than 0.05\,\% in magnitude.}
\end{table*}

The agreement between the simulation and the data is quite good, especially for all of the data requiring a photon coincidence ({\it i.e.}, the plots in the second and third columns of Figure~\ref{fig:MC_Data}). The good agreement in the proton arrival time plots indicates that the tracking algorithm and the field simulations perform well. The overall agreement gives us confidence in the simulation and our understanding of the apparatus, but it is important to note that exact reproduction of these data is not necessary to obtain the correct branching ratio and photon spectra at a level on the order of 1\,\%.

\section{Systematic corrections and uncertainties}
\label{sec:systematics}

This section discusses the systematic corrections and uncertainties associated with measuring the radiative photon energy and branching ratio. The discussion includes systematic effects associated with the cuts in Table~\ref{tab:cuts}. The uncertainty for each effect was assigned by varying the range of each cut to determine the resulting change in the ratio $R$. Simulation-related systematics are also discussed. Because there are two distinct photon detectors that measure two energy regions, their systematic effects are addressed separately.  Table~\ref{table:Systematics} summarizes the systematic corrections and uncertainties.

\subsection{Photon detectors}

\subsubsection{Photon energy calibration}
\label{sys:photoncal}
\paragraph{\textbf{BGO}}
\label{sys:photoncalBGO}

\begin{figure}
 \includegraphics[width=0.48\textwidth]{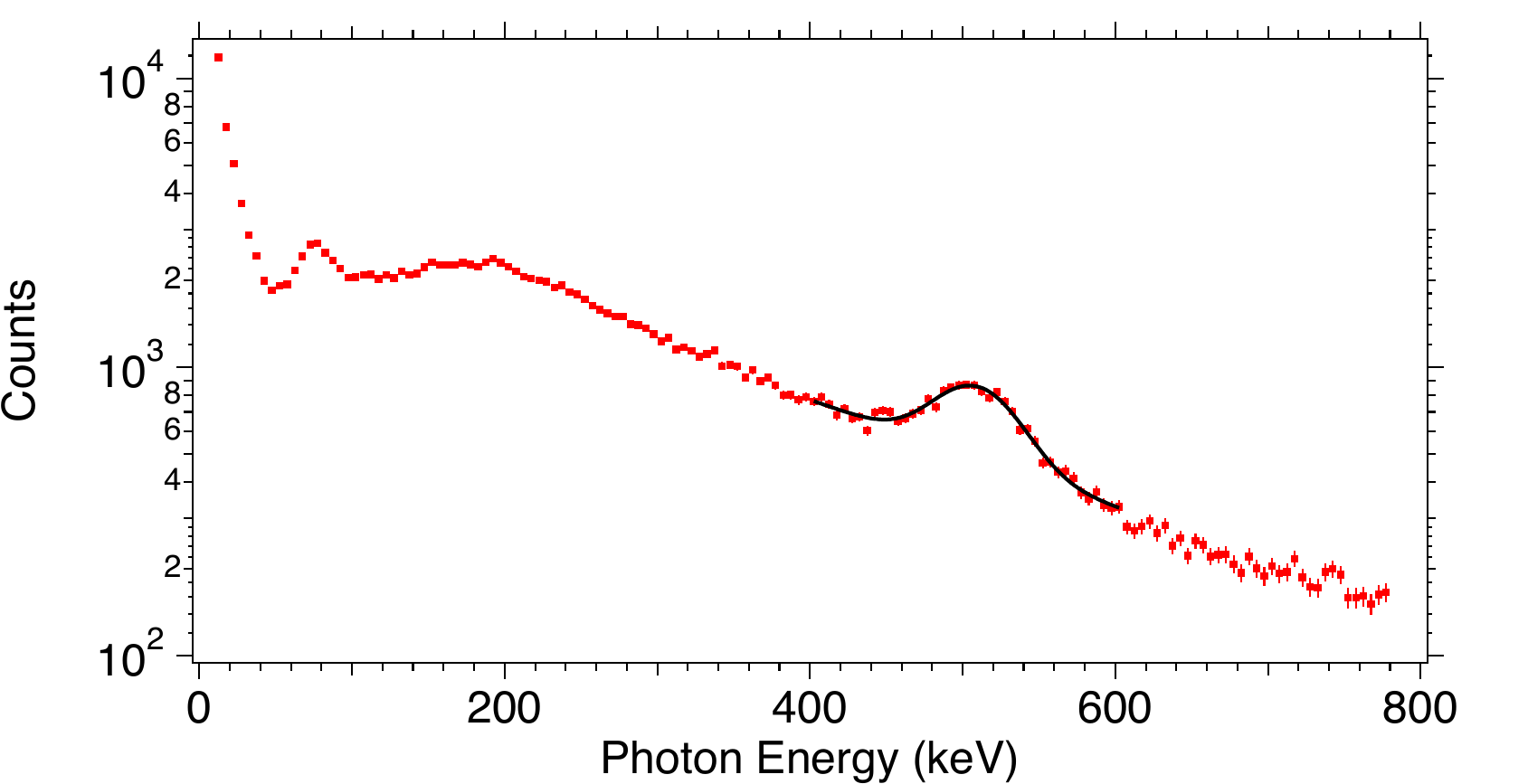}
 \includegraphics[width=0.48\textwidth]{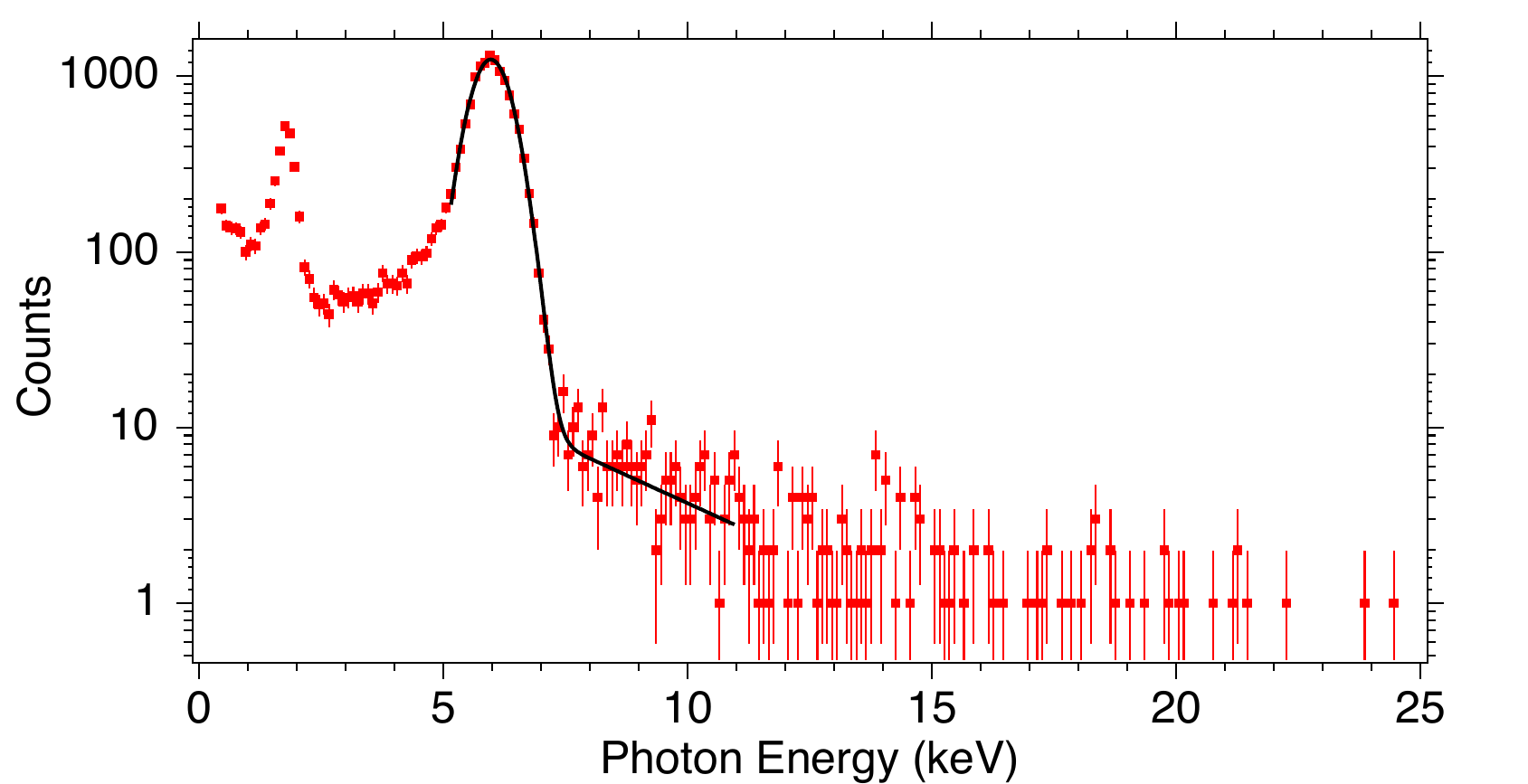}
 \caption {\label{fig:DetCal} (Top) The energy spectrum of the full BGO data set, prior to background subtraction. Each BGO detector was calibrated and then the 12 spectra were summed to generate this spectrum. The broad feature centered around 200\,keV is largely from the Compton scattering of high energy photons, and the peak around 80\,keV arises due to the escape of bismuth K-shell x-rays from nearby crystals. The black line is a fit to the 511\,keV line using Eq.~(\ref{eqn:photoncal}). (Bottom) The energy spectrum of the full APD data set, prior to background subtraction. Each APD detector was calibrated and then the 3 spectra were summed. The black line is a fit to the 5.9\,keV x-ray from \Iff. The error bars represent the statistical uncertainties only.}
 \end{figure}

Calibrations and studies of the BGO and APD detectors were performed both online and offline. During online data acquisition, the DAQ could be placed in a mode to collect calibration data.  At the beginning of each run and every twelve hours thereafter, calibration data were acquired for 30 minutes while the beam shutter was open. In this mode, the DAQ triggered on all detectors rather than only after a valid electron-delayed proton event. This mode produced a spectrum of photons in which the 511\,keV from electron-positron annihilation introduced an easily identifiable peak. It was observed in the photon background spectrum of all the BGO detectors and was used as a continuous monitor of the calibration.
The 511\,keV peak fit well to a Gaussian distribution on an exponentially decaying background with the function
\begin{equation}\label{eqn:photoncal}
   \mathcal{G}(E) = a + be^{-\lambda E} + ce^{-(\frac{E-E_\circ}{\sigma})^2},
\end{equation}
\noindent where $a$, $b$, and $c$ are constants of the fit, $\lambda$ characterizes the exponentially decaying background, $E_\circ$ is the position of the 511\,keV peak, and $\sigma$ parameterizes its width.  Figure~\ref{fig:DetCal} shows the sum of all the calibrated energy spectra used in the analysis.

After online data collection was complete, calibrations were conducted offline using $^{137}$Cs, $^{57}$Co, $^{133}$Ba, and $^{241}$Am sources. These runs were used to refine the response functions used in the Monte Carlo simulations. For the runs, a thin-wall steel tube wrapped
in insulating aluminized mylar was placed along the center of the detector. This allowed data to be collected from radioactive samples placed at positions along the center of the detector while the apparatus was cold. A typical BGO detector energy resolution was 10\,\% (full width at half maximum) at 662\,keV and  30\,\% at 60\,keV~\cite{Cooper2012}.

The dominant source of calibration uncertainty comes from the determination of the shape and offset of the baseline fitting the photon waveforms. To estimate the size of the effect, the data were fit to Eq.~(\ref{eqn:template}) allowing for both a flat background and a sloping background. The differences in the pulse heights from the two methods were histogrammed, yielding a 0.16\,keV shift at all energies. That value produces a change in $R$ (the ratio of the \epg\ and \ep\ rates) of 0.6\,\%, which is assigned as the calibration uncertainty for the BGO detectors. No correction was assigned to this systematic as allowing for a slope in the waveform fit was considered to be the more accurate approach and any residual uncertainty is assumed to be included in the larger uncertainty associated with the energy response.

\paragraph{\textbf{APD}}

For the APD detectors, the risetime of the preamplifier tail pulses was much faster than the signals from the BGO scintillator, as seen in the widths of the two timing spectra in Fig.~\ref{fig:egTime}. The APD pulse height was obtained from the maximum of the tail pulse after de-noising and a linear fit to the waveform baseline. During data collection, the APDs were exposed to a weak \Iff\ radioactive source mounted near the detectors, which produced 5.9\,keV photons for a continuous calibration.  Offline studies at synchrotron sources were performed to explore the complex energy response of the APD, which is due to the reduced charge collection efficiency for photons that are absorbed in the front 1\,$\mu$m of the APD~\cite{Gentile2012}.  Models of the charge-collection efficiency of the APDs' doped layer of Si were created and incorporated into the MC simulation of the detectors.

Calibration factors were obtained for each detector in each series and applied to the pulse height spectra. The pulse spectra were histogrammed, and the data around the Fe-55 peak were fit to a weighted Gaussian and an exponential background. The peak value was used to calibrate each detector in each series. Figure~\ref{fig:DetCal} shows the sum of all the calibrated energy spectra for the electron-photon timing region used in the analysis. The peak at 1.7\,keV is an intermittent artifact associated with the high voltage on the SBD. It is assumed to be a silicon escape peak, but it was not investigated in detail. It contributed to the random background and subtracts out in the analysis.

The dominant source of uncertainty in the calibration comes from using the peak height determination rather than a fit to the waveforms. It does not produce a multiplicative change to the calibration factor but manifests itself as an offset. The peak height determination typically selected a value that was too high because of noise on the waveform. To estimate the magnitude of this effect, a series was selected, and all of the APD waveforms were fit to a sigmoid function with an exponential to account for the fall time of the tail pulse. We then fit the background to a line and subtracted the value of the background at the maximum value of the sigmoid fit.

From the difference between this fit method and our standard peak-height method, an average offset of 0.05\,keV was determined for the events. If one assumes that this generates an energy offset common to all detectors, one can use it to estimate a calibration uncertainty for the data set. One could correct all the data points by 0.05\,keV, or one could just remove that fraction of the events from the first bin. We chose to do the latter and determined the uncertainty to be 11 out of 849 events, corresponding to 1.3\,\%.

\subsubsection{Energy response}
\label{sys:enresponse}

The uncertainty due to the energy response of the BGO detector was dominated by uncertainty in the nonproportional response of BGO~\cite{Khodyuk2012, Moszynski2004, Verdier2011, Sysoeva1996, Averkiev1990}. Nonproportionality refers to the phenomenon where a scintillator's light output is not directly proportional to the energy deposited.
The nonproportionality was modeled by applying the results of a parametrization given by $N_E = a/(1+bE)$, where $N_E$ is the nonproportionality for an energy deposit $E$, and $a$ and $b$ are fitted parameters.   Although this form was purely phenomenological and not based on the physics of BGO nonproportionality, it provided a convenient functional form that matched the data well.  The variation of nonproportionality with energy shows features that are correlated to the bismuth atomic structure. Hence, this form was fit for four energy regions: 90\,keV to 1000\,keV (above the K edge), 16.5\,keV to 90\,keV (between the L and K edges), 13.4\,keV to 16.5\,keV (the L edge region) and below 13.4\,keV (below the L edge region). Data for nonproportionality, both from our own studies and prior reports, are summarized in Ref.~\cite{Gentile2015}. 

\begin{figure}[h]
 \includegraphics[width=0.48\textwidth]{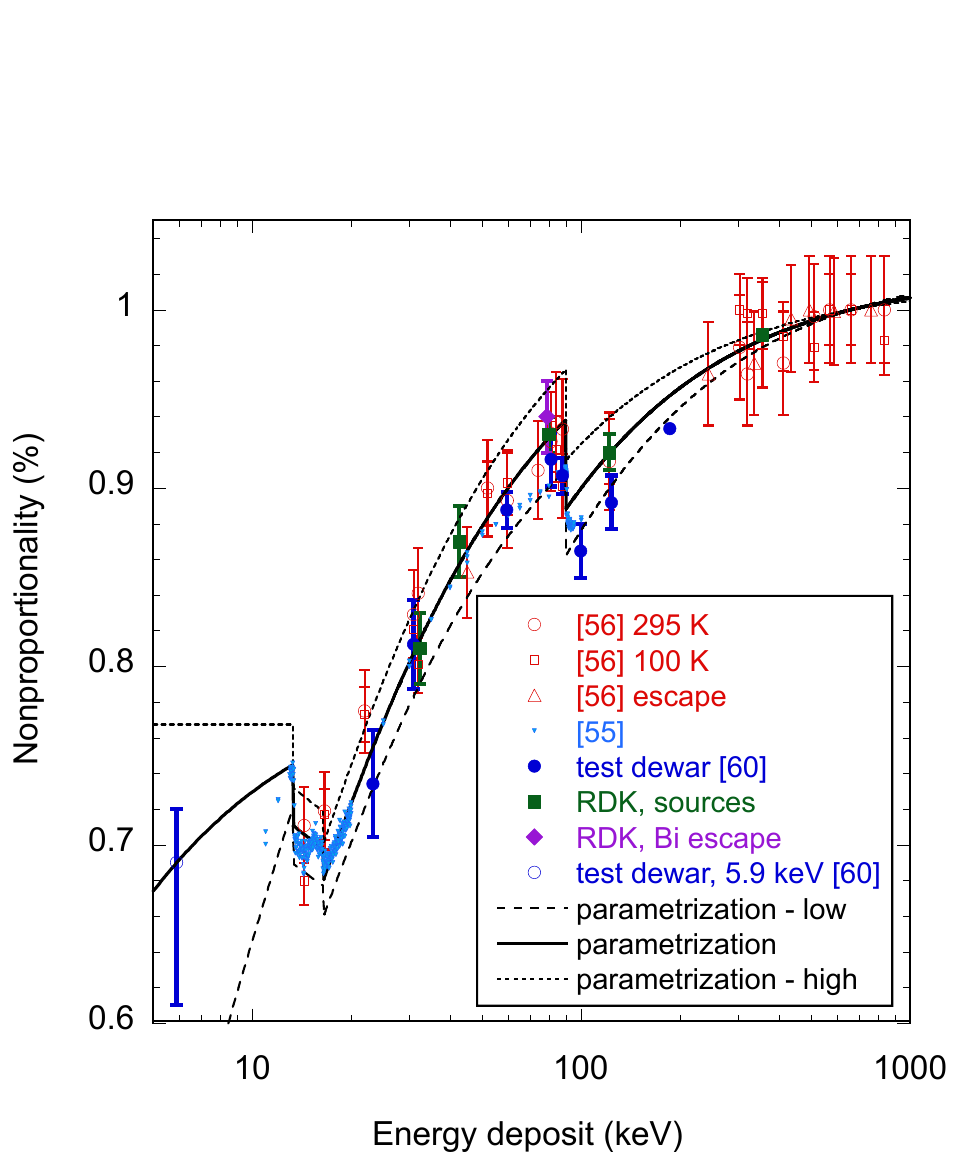}  
 \caption{Parametrization employed for BGO nonproportionality, which is based on the data sets in the plot. The form is discussed in the text.  The curves labeled ``high'' and ``low'' correspond to variations of the parametrization that yielded a 2.6\,\% uncertainty in the branching ratio due to energy response. The error bars shown are those reported in the reference listed or our estimated uncertainties for measurements taken in the RDK II apparatus.}
 \label{fig:nonpropwithmodelsFe} 
 \end{figure}

Figure~\ref{fig:nonpropwithmodelsFe} shows data from the literature~\cite{Moszynski2004,Khodyuk2012}, our test dewar, the RDK~II apparatus, and our parametrization.  For the region below 13.4\,keV, only a small subset of data from Ref.~\cite{Khodyuk2012} and our test dewar result at 5.9\,keV were available. Prior experiments (see Ref.~\cite{Gentile2015}) had shown a range of results, including nonproportionality actually increasing with decreasing energy.  We chose a curve that passed through our 5.9\,keV datum. 

To assign an uncertainty in the branching ratio due to the uncertainty in our nonproportionality model, we varied the parametrization to accommodate the range of values obtained in the four data sets. For the two regions between 16.5\,keV and 90\,keV, we varied $N_E$ by $\pm$3\,\%. For the region above 90\,keV, we varied $N_E$ at 90\,keV by $\pm$3\,\% but constrained it to be unity at 662\,keV.  For the region below 13.4\,keV, the ``high'' curve was simply flat, while the ``low'' curve passed through the value of $N_E$ that would have been obtained from the location of our observed peak for 5.9\,keV without the asymmetric fit employed in Ref.~\cite{Gentile2015}. Because we only analyze radiative decay data above 14.1\,keV, nonproportionality below this energy was only relevant because of the increased broadening at low photon energies. The energy resolution, which is dominated by photoelectron statistics and thus varies as the inverse of the square root of photon energy, is 60\,\% at 14.1\,keV.   The generous range for possible nonproportionality at low energy had little effect on our total uncertainty associated with nonproportionality. The effect of using a Gamma distribution instead of Gaussian broadening resulted in less than 0.1\,\% effect on this value. Another small systematic effect was due to the $\approx$10\,\% variation in light output along the BGO crystal~\cite{Cooper2012}.  The measured curves from our calibration measurements were incorporated into the simulation directly, and a 0.3\,\% effect on the branching ratio uncertainty was determined by turning this on and off in the simulation. The total uncertainty in the branching ratio was 2.6\,\% and primarily due to nonproportionality.

Because the APD energy response is complex, uncertainty arose from several sources.  In Ref.~\cite{Gentile2012}, a model for the dependence of the electron collection efficiency on the depth at which x-rays were absorbed was determined from measurements performed using monochromatic x-ray beams. It was tested by comparing a prediction for the response to broadband synchrotron radiation to experimental data.  The optimization was done by adjusting the parameters by hand to yield the best reproduction of the monochromatic beam data.  The model could also be optimized for the broadband case; hence, an uncertainty could be estimated by evaluating the effect on the branching ratio for the two approaches.  For the one APD that was tested with both monochromatic and broadband x-rays, the simulated branching ratio was found to shift by about 5\,\%.  The broadband-based optimization was required for the other two APDs employed because they were not studied in the monochromatic beam experiments.  Whereas one of these APDs had very similar characteristics, the third APD that had been purchased later was found to yield better low-energy response but also exhibited some deviation from the expected proportionality between pulse height and x-ray energy~\cite{Gentile2012}.  This APD was well-modeled using the broadband data, but the absence of monochromatic beam data for comparison yields additional uncertainty.  In addition, the nonproportionality was bias voltage dependent, and the off-line tests were not known to be at exactly the same voltage below breakdown as for online operation of the experiment.  

Another source of uncertainty was due to the strong magnetic field employed in the radiative decay experiment, as all of the APD characterizations were done at essentially zero field.  We observed that the resolution improved by almost a factor of two at high field~\cite{Gentile2011}.  This narrowing has only a small effect on the simulated branching ratio, but the observation of such an effect from the magnetic field indicates that there may be other small changes in the APD response.  Combining all of these effects, we estimated an uncertainty in the branching ratio of 10~\% due to the energy response of the APD.

\subsubsection{Multiple photons}
\label{sys:multiplephotons}

In the photon detectors, there are the true coincident \epg\ events but also \epg\ events where the photon is from a background that is independent of the neutron decay process. In some instances,  both types of events happen by random coincidence in the same BGO trace producing pulse pileup.  Systematic errors can be introduced by both adding mistaken events and removing legitimate events. The events are rare due to the low single-photon event rates in the detectors, but they are sufficiently frequent that one must address how those random $\gamma$-$\gamma$ coincidences affect the result for the branching ratios in both detectors.

The multiple photon effect scales linearly with background count rate in each BGO detector for a given background energy spectrum. The average background rate is ($28 \pm 10$)/s in each detector for photons between 10\,keV and 1\,MeV. There are roughly 20\,\% more background events with energies $>1$\,MeV (i.e., signals that saturate in the preamplifier) and a small number of discarded fast risetime events (e.g., electronic noise or direct absorption of gammas in the APD). Because a trace with two signals results in only one event emerging from the analysis, to a good approximation there is a net loss of one \epg\ event for each trace showing the coincidence of a true and random event. It produces a loss of 0.2\,\% of \epg\ events.

There is a similar loss of \epg\ events (about 0.13\,\%) for traces where a large-energy random signal arrived before the beginning of the trace, but the real \epg\ signal occurred on the tail of the undetected random signal. This loss varies with \epg\ energy, and for example, the amounts are about 0.24\,\% at 15\,keV and 0.05\,\% at 85\,keV. Overall, there was a loss of $(0.4 \pm 0.2)$\,\% of \epg\ events for random background gamma-rays. If  both signals fall within or near the \epg\ coincidence window the standard random subtraction for those events subtracts the wrong energy spectrum, but those affect less than 5 events out of 20,000 \epg\ events. Finding two randoms on a trace (instead of one random and one coincident \epg\ gamma) has no net effect on the \epg\ extraction.

The analysis for the low-energy photons detected by the APDs follows similarly. We considered two-photon events on a single trace and one photon on the digitizer trace with a second photon starting before the recorded trace, but with a significant part of the signal tail present on the trace. The investigation found that ($0.1\pm 0.1$)\,\% of the coincident \epg\ APD signals were lost due to pileup. The reasons for the smaller effect of pileup in the bare APDs compared to the BGO crystals are (a) the count rate of background is smaller (about 14/s per detector) and (b) the background energy spectrum compared to the \epg\ gamma spectrum is softer than for the BGO data with very few counts above the 5.6\,keV Fe calibration peak, thus very few large amplitude pulses that result in an extended tail. The corrections were applied for both the BGO and APD data.

\subsection{Silicon detector}

\subsubsection{Electron and proton energy calibration}
\label{sys:SBDCal}

Because the branching ratio depends on Monte Carlo simulation, the electron and proton calibrations are important and do not drop out in the ratio. The electrons and protons were detected by the same SBD, and thus their calibrations were determined in the same manner. SBDs are sensitive to gamma rays, and we presume a response for a given energy gamma to be the same as the response from a given energy charged particle. We also assume linearity in both cases. The SBD calibration came from two methods: fitting the endpoint of the beta spectrum and calibrating the detector with \Atfo\ and \Cofs. The exercises described in this section quantify uncertainties associated with calibrating these detectors.

To obtain a calibration factor from the endpoint, the standard $ep$ cuts were made to all the data in the series, and the resulting spectra were fit for the beta endpoint with the functional form
\begin{multline}\label{eqn:endpt}
    \mathcal{H}(E) = B + A\sigma \frac{(E_\circ-E)}{\sqrt{2\pi}}e^{-\frac{(E_\circ-E)^2}{2\sigma^2}} \\
 + \frac{1}{2}(\sigma^2+(E_\circ-E)^2)\left\{1+\erf\left[\frac{(E_\circ-E)}{\sqrt{2}\sigma}\right]\right\},
\end{multline}
\noindent where $E_\circ$ is the endpoint value, $\sigma$ is the experimental resolution, $A$ is an amplitude, $B$ is a constant background, and where $E$ is the peak height, which is proportional to energy for an electron, proton, or gamma ray. This equation was obtained by convolving the dominant behavior at the endpoint with a Gaussian response function. It was multiplied by the remaining terms of the function describing the spectral shape, which would only be slightly affected by the convolution.

\begin{figure}[h]
 \includegraphics[width=0.48\textwidth]{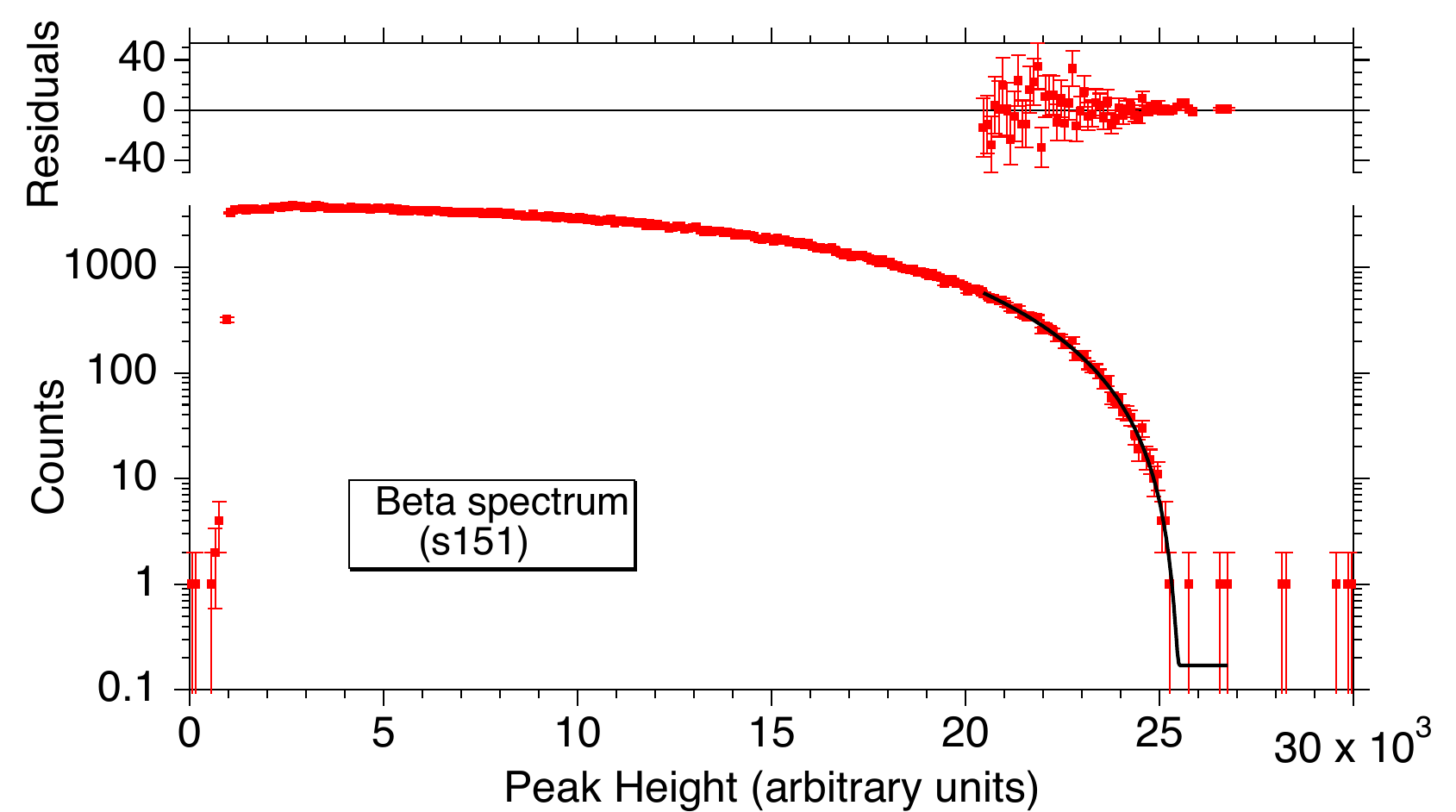}  
 \includegraphics[width=0.48\textwidth]{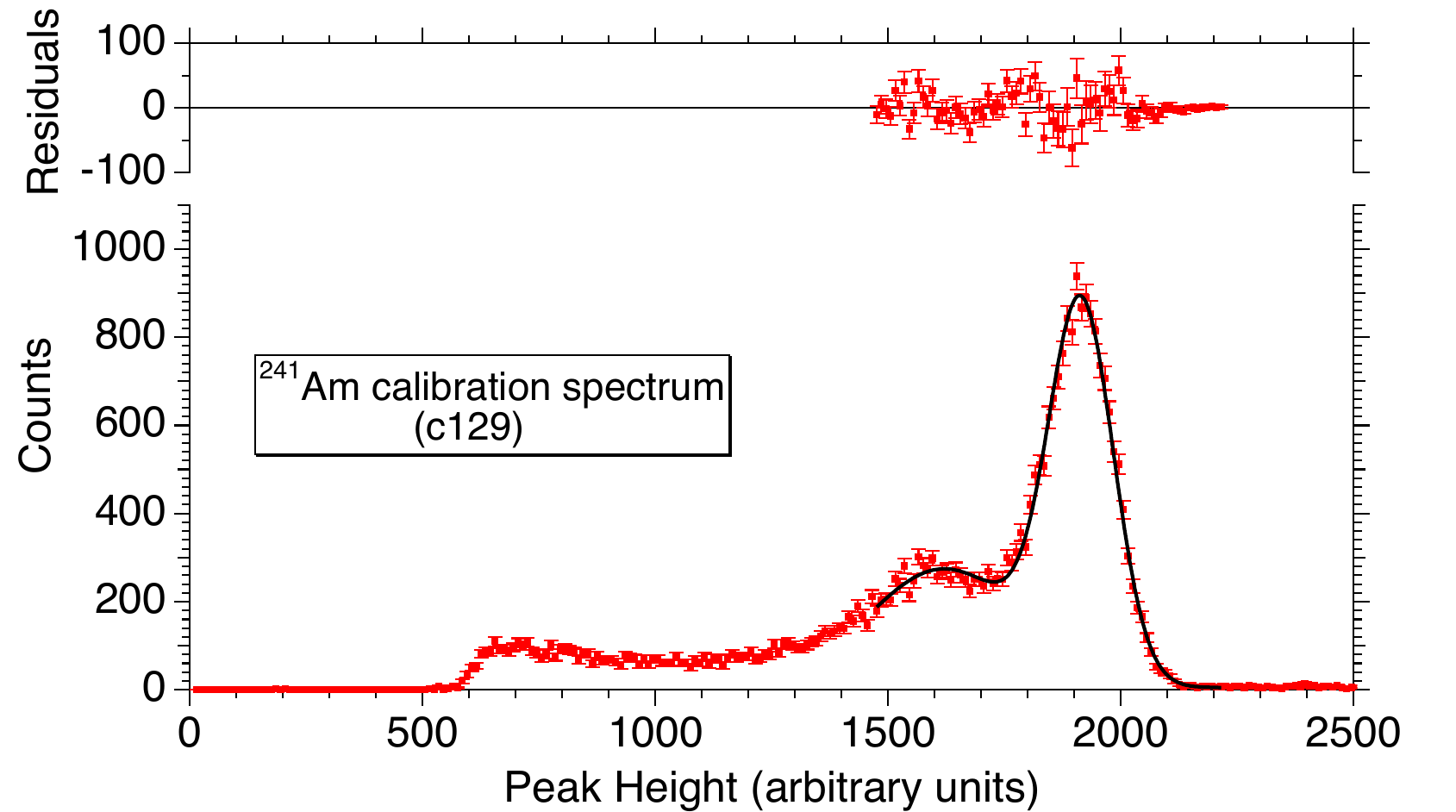}  
 \caption {(Top) Fit to the beta spectrum near the endpoint for an arbitrary series. The fit region was restricted to reduce the influence of the low energy part of the spectrum. (Bottom) Fit to the 59.5\,keV x-ray from an external \Atfo\ source. The error bars represent the statistical uncertainties only.}
 \label{fig:SBDCals} 
 \end{figure}

Both $\sigma$ and $B$ were held constant in the fit. The resolution was determined to be 2.3\,keV, and the background was determined by averaging the counts above the endpoint. Figure~\ref{fig:SBDCals} (top) gives an example of a fit to the electron spectrum for a series. The offset in the amplitude was neglible in this context, and a two-point calibration was performed using the neutron beta-decay end point energy of 781.58\,keV and 0, taking into account that the SBD floated at $-25$\,keV. The resulting calibration factor was applied to all electron and proton events in the series; each series had its own calibration factor.

Calibrations using external \Atfo\ and \Cofs\ sources were performed about once per reactor cycle, and they were used as a check on the endpoint calibrations.  For \Atfo\, a region near the 59.5\,keV peak was fit to two Gaussians and a constant background to obtain the peak position (see Fig.~\ref{fig:SBDCals} (bottom)). The analysis for \Cofs\ calibrations were done similarly using the 122\,keV gamma peak. The endpoint method was chosen as the primary method because it provided a continuous monitor of the spectrum for all the series and the \Atfo\ and \Cofs\ calibrations were performed only periodically.

There are two dominant ways in which the calibration can contribute to uncertainty in the position of the energy cut. One way was if there were an error in the calibration, and hence the cut was at the wrong value, and another way was if the detector resolution moved events across the threshold. The latter effect is negligible because the number of events around the cut values is comparatively flat, resulting in little net loss or gain of events. 

The absolute calibration, however, does produce a small uncertainty. There is an absolute difference in energy from the two methods at the lower cut value for the electrons and protons. The difference in the two calibration methods was used as a measure of the calibration uncertainty. The assessment of the sensitivity of the ratio $R$ to the cut is done in the same manner as that of the other systematics. One calculates $R$ as the position of the cuts varies around the standard cut values. The standard cut value at the electron low-energy threshold was 100\,keV, and we took the average difference in energy of the two calibration methods of 2.2\,keV. This yields a correction of $0 \pm 0.2$\,\% for the BGO ratio and $0 \pm 0.5$\,\%  for the APD ratio. The upper cut value contributed negligibly because there are so few events at the electron endpoint. 

The same procedure was followed for the systematic uncertainty from the calibration for the proton cuts. The acceptance range was 13\,keV to 31\,keV. Here, both values are relevant because there are substantial numbers of events at each energy. The average difference in the two calibrations is 0.9\,keV. That value was varied at the lower and upper limits to determine the magnitude of the change in $R$. It yields a correction of $0 \pm 0.5$\,\%  for the BGO ratio and  $0 \pm 0.4$\,\%  for the APD ratio.

\subsubsection{Waveform identification}
\label{sys:WaveformID}

\begin{figure}
\includegraphics[width=0.48\textwidth]{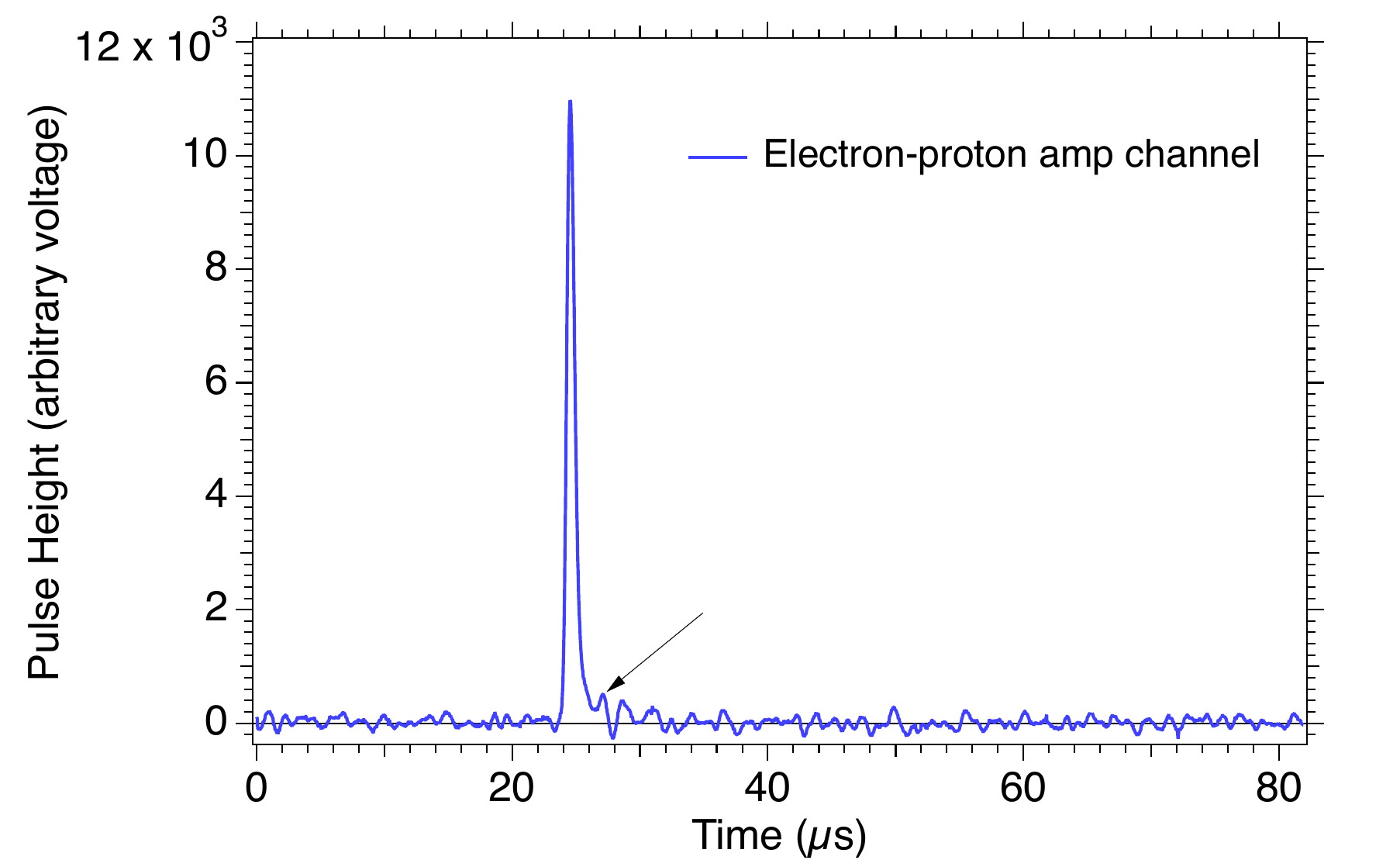}
\includegraphics[width=0.48\textwidth]{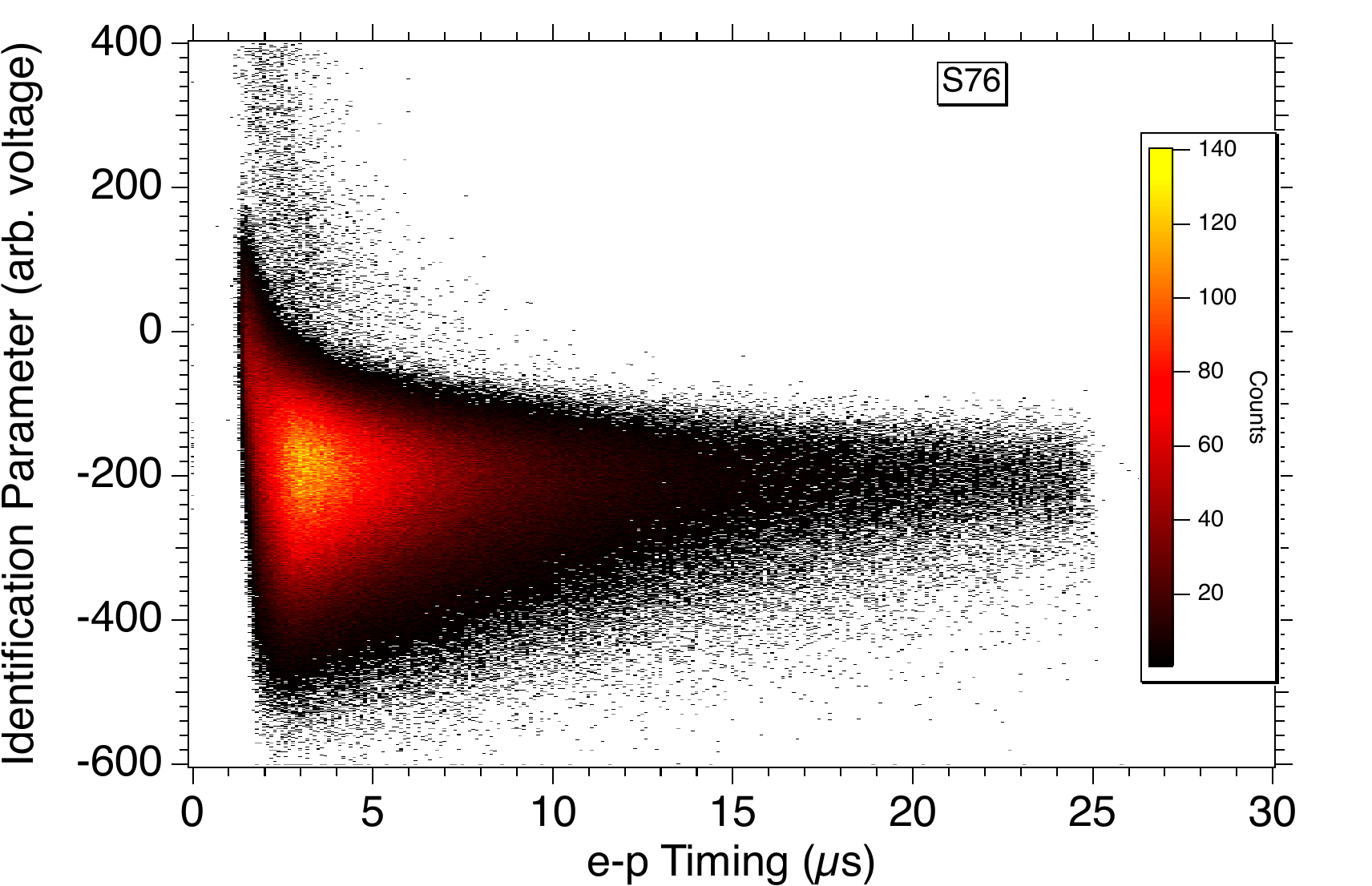}
\caption{\label{fig:CooperEvent} (Top) This event triggered the DAQ to record it as a candidate electron followed by a delayed proton event. There is ambiguity as to whether or not the DAQ was triggered by a candidate proton or if it was noise on the electron tail that was above the trigger threshold (see the arrow in the plot). (Bottom) 2-D histogram of waveform identification parameters versus the $ep$ timing. The waveform cut value is the minimum value of the trace between the arrival time of the electron and the proton.}
\end{figure}

This section addresses the uncertainty in the branching ratio that arises in identifying proton events that arrive close to the electron pulse in time. Figure~\ref{fig:CooperEvent} gives an example of a waveform for which the identification of the proton event is not as obvious as in the example of Fig.~\ref{fig:waveforms}.
One must differentiate true protons from noise on the tail of the electron pulse that triggered the DAQ. There are two parameters that are relevant in making a distinction: the time after an electron event at which one starts to look for a proton event and the minimum value of the trace voltage between the arrival times of the electron and the proton, which is referred to as the waveform identification parameter.

The bottom plot of Fig.~\ref{fig:CooperEvent} shows a 2-D histogram of the distribution of the waveform identification values versus the $ep$-timing value for a given series. An optimization was performed to establish the values to use for both the cuts in the standard analysis, but there is some arbitrariness in the selection of the values. The $ep$-timing cut was selected to be 2\,$\mu$s. If it were shorter, the proton could be lost in the electron event, and if it were longer, one starts discarding too many proton events. The waveform identification cut was selected to be 100 in arbitrary units of digitizer voltage, {\it i.e.}, the waveform was required to go below 100 between the electron and the following peak that passed the proton trigger.

To understand how the waveform cut affected the ratio $R$, the value was varied from $+200$ to $-200$ and also included no cut at all. We also examined the same distribution for other series to ensure that there was no variation over the duration of the experiment. The analysis and Fig.~\ref{fig:CooperEvent} indicate that a value of 200 is the most conservative, and a value of 0 is too restrictive. The value of 100 was selected as a reasonable compromise between accepting invalid events and discarding real events. 

For the uncertainty associated with this cut, we adopted a conservative approach of using half of the range in $R$ when varying the cut value from 0 to no cut at all. For the BGO data, it yields an uncertainty on $R$ of $\pm 2.2$\,\%, and for the APD data, the uncertainty is $\pm 0.4$\,\%. There is a much weaker dependence of $R$ on the waveform cut for the APD data, but we do not ascribe any particular significance to that difference.

\subsection{Timing cuts}
This section addresses the systematic uncertainties arising from the selection of the $ep$ timing cuts and the timing region used for the $e\gamma$ windows.

\subsubsection{Electron-proton timing}
\label{sys:eptiming}

 The $ep$ timing cut is from 2\,$\mu$s to 25\,$\mu$s, or 50 to 625 in digitizer channels. For the conversion to time, the board digitization of one channel corresponding to 1/(25\,MHz) ({\it i.e.,} 40\,ns) was used. To quantify the uncertainty arising from the cut, one must understand the uncertainty in determining the time difference between the electron and proton. As discussed in Section~\ref{subsec:DAQ}, the $e$ and $p$ waveforms from the shaping amplifier were fit to a Gaussian function with a linear and linear plus an exponential background term, respectively, to obtain the position of the peak channel. The timing is extracted from the position of the Gaussian peak, and each event is assigned a $\Delta_{ep}$ time value, i.e., the arrival time of a proton after the prompt detection of an electron.
 
\begin{figure}
\includegraphics[width=0.48\textwidth]{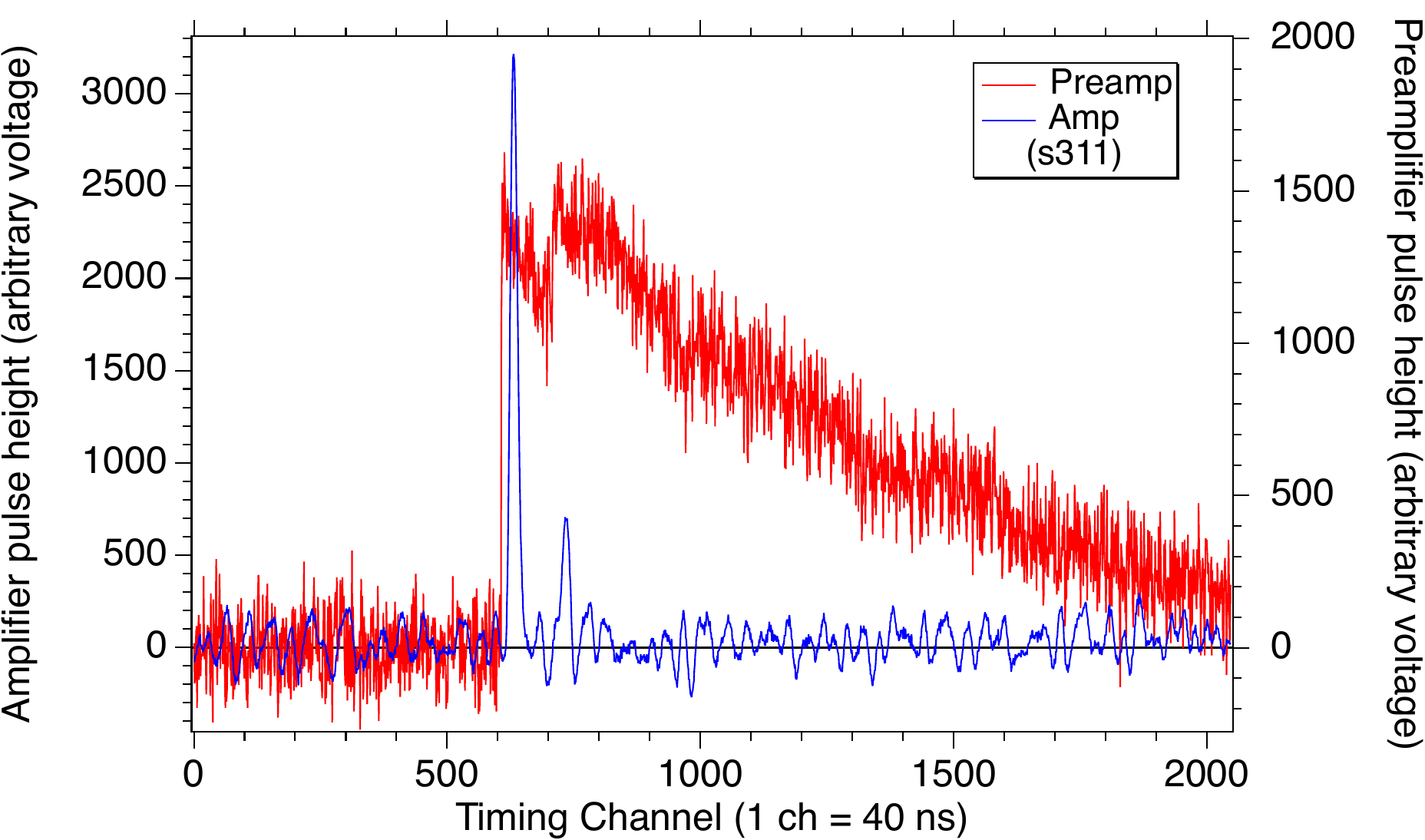}
\caption{\label{fig:TimingCut} Plot showing an example of an event for extraction of the proton arrival time relative to an electron event.}
\end{figure}

Figure~\ref{fig:TimingCut} shows a typical $ep$ event in both the preamp and shaping amp channels. The $ep$ time difference is the onset of the proton pulse minus the onset of the electron pulse in the preamplifier waveform, but we used the shaped pulse from the amplifier to get the time differences. The shaping amplifier integrates (250\,ns) and differentiates the preamp pulse, so the peaks of the $e$ and $p$ pulses correspond to the inflection points of the preamplifier pulses.

The uncertainty in the timing position from the fit is small, on the order of a fraction of a channel, but the larger contribution comes from the noise on the waveform. It is particularly noticeable when fitting proton events because their amplitude is not that much larger than the noise amplitude. A simulation was performed to quantify the size of the effect from noise. Gaussian noise was added to an ideal proton waveform. 5000 timing spectra were generated, and the fitted peak position and its standard deviation were tallied. This was done for four different pulse heights, selected to span the typical proton pulse amplitudes of 300, 400, 600, and 800 in arbitrary voltage units. The standard deviations ranged from 0.9 channels for the highest amplitude to 2.5 channels for the lowest amplitude. From this analysis, a conservative value of 3 channels was taken to be the uncertainty in the timing.

To determine how sensitive the ratio $R$ is to the cut, the $ep\gamma/ep$ ratio was evaluated while varying the timing window by $\pm 3$ channels (120\,ns). For the BGO data, varying the timing window by that amount gives an uncertainty of $\pm 0.5$\,\%, and for the APD data one obtains an uncertainty of $\pm  0.6$\%, comparable to the value for BGO. There is no correction assigned with this systematic uncertainty, and there is a negligible contribution from the cut value of channel 625 at the end of the window due to the small amount of data there.

\subsubsection{Electron-photon timing}
\label{sys:egtiming}

Figure~\ref{fig:egTime} shows the histogram for the time difference between the electron and photon events for both the BGO and APD detectors. There are three relevant regions: the prepeak background, the peak, and the postpeak background. The two background regions were weighted by their size to obtain the background correction. The background correction was essential for determining the value of the branching ratios and also to obtain the energy spectra. The systematic effect that is of concern is the possibility that there are true radiative decay events occurring outside the window.  The uncertainty can be made arbitrarily small by choosing a sufficiently large window on the peak, but one aims to make the window as small as possible to reduce statistical fluctuations associated with the background subtraction while not excluding real events.

It is straightforward to check by changing the size of the peak window and evaluating the effect on the branching ratios. The size of the window was varied between 4\,$\mu$s and 8\,$\mu$s. (The endpoints of both the prepeak and postpeak background regions stay the same, meaning as the size of the peak window decreases/increases, the size of the background regions increases/decreases correspondingly.) For the BGO data, the ratio $R$ stabilizes with a window width of $\geq 6\,\mu$s, which is what was selected for the analysis. When the window was opened to 8\,$\mu$s, the ratio increased by less than 0.1\,\%. This is a negligible change in both the value and the error bar. As such, no correction was assigned for this systematic effect and the uncertainty was considered to be negligible.

The same analysis was performed for the APD detectors with a similar result. A window that is $\geq 0.6\,\mu$s wide was sufficient to contain the photon events with a negligible loss. Recall that the timing peak is much narrower than the corresponding one for BGO detectors due to the faster rise time of the APD signal (see Fig.~\ref{fig:egTime}). In addition to this analysis of varying the window size, we also fit the $e\gamma$ timing spectrum to a Gaussian and used the fitted parameters to determine the number of counts outside of our acceptance window. The background was determined using the entire background region and then held as a constant for the fit. As in the analysis performed by varying the window size, there was a negligible contribution from the Gaussian tails outside of the acceptance window. This analysis was not done on the BGO $e\gamma$ spectrum due to the non-Gaussian shape of the peak.

\subsection{Correlated $ep\gamma$ backgrounds}
Uncorrelated backgrounds are subtracted out in the timing spectrum, but correlated backgrounds must be addressed separately. Two such sources of correlated background, true correlated events from electron bremsstrahlung and electronic artifacts, are discussed in this section.

\subsubsection{Electron bremsstrahlung}
\label{sys:brems}

The bremsstrahlung of electrons that lose energy in the SBD could strike a BGO crystal and induce an $ep\gamma$ signal that is indistinguishable in the analysis from a true radiative decay event. Such events were rare due to the geometry of the detectors; the BGO and APD presented a small solid angle from the SBD and were shielded where possible (see Fig.~\ref{fig:diagram}).

The MC simulation was necessary to determine the number of events that would be erroneously attributed to radiative decay in the data set. This was accomplished by running a three-body decay in the simulation and tallying the number of bremsstrahlung photons produced by the electrons in the SBD that were subsequently detected in the BGO or APD arrays. The MC was run with the Penelope library with 200 million decays, producing approximately 20 million $ep$ coincidences. Of those 20 million, there were no energy deposits in the APDs and 112 responses total in the twelve BGO crystals in the energy range of 10\,keV to the end point. The lack of any events in the APDs is not surprising, given that the low-energy photons were unlikely to penetrate the mounting hardware. However, the BGO detectors did have some direct visibility to the SBD.

\begin{figure}[t]
\includegraphics[width=0.48\textwidth]{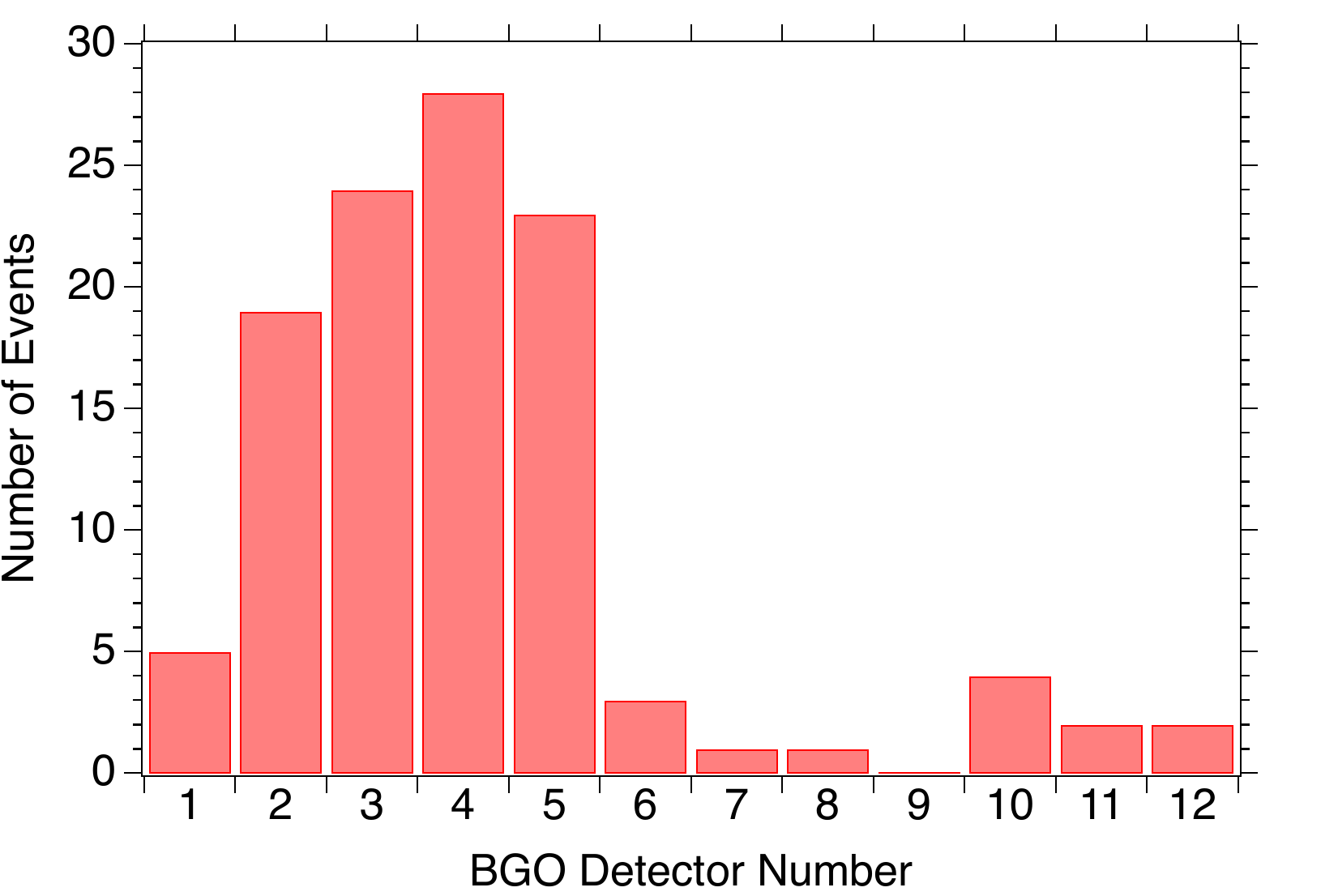}
\caption{\label{fig:Brems} Plot of the MC distribution of bremsstrahlung events in the BGO detector array.}
\end{figure}

Figure~\ref{fig:Brems} shows the distribution of bremsstrahlung photons in the BGO detector crystals. The MC indicates that the majority of events would be expected in the crystals mounted away from the SBD. Looking at the arrangement of the BGO array relative to the SBD, the distribution is expected because there is a greater line-of-sight to the SBD for those detector crystals (see Figs.~\ref{fig:diagram} and \ref{fig:Detmap}). Tallying the bremsstrahlung photon ratio and true radiative event ratio gives $R_{\rm brem} = (6.6\pm 0.9)\times10^{-7}$ and $R = (819\pm 3.7)\times10^{-7}$, respectively. This results in a $-0.8$\,\% correction and a $0.14$\,\% uncertainty in the BGO detectors and no detected photons in the APDs.  All of the uncertainty comes from the MC statistics.

\subsubsection{Non-decay backgrounds}
\label{sys:nondkbkgd}

In addition to electron bremsstrahlung causing a correlated background in the timing spectrum, it is also possible for correlated events unrelated to radiative neutron decay to survive all of the cuts and appear in the timing spectrum. The origin of any given correlated event is difficult to discern, but it is believed that they can arise from cosmic events, secondary cascades from (n,$\gamma$) capture events, and electronic artifacts or glitches. Although the number of such events is not large, as most are rejected by the cuts, they do occur with sufficient frequency to warrant investigation.

Correlated non-decay background could be measured directly if one were able to ``turn off'' neutron decay while keeping all of the other experimental parameters the same. Although this is not a possibility, we can remove our sensitivity to true radiative decay events by making measurements with 1) the high voltage on the SBD at zero volts, 2) the magnet field (B) off, and 3) the neutron beam off. Coincidences from these runs will be dominated by non-decay backgrounds, but they still have weaknesses: the conditions are not completely realistic; the datasets are comparatively short, so the statistics are poor and there is little energy information; and the checks were made during early production running only.

Regardless, each of these runs should not produce correlated neutron decay events in the timing spectrum, so quantifying their occurrence gives information on their contamination of the true radiative signal. The standard set of cuts were applied to these data yielding:

\begin{enumerate}

\item[1)] HV = 0 (S97): 6 events/0.93\,d = 6.4/d
\item[2a)] B=0 (S98): 5.2 events/0.95\,d = 5.5/d
\item[2b)] B=0 (S99): 2.8 events/0.50\,d = 5.6/d
\item[3)] beam off (S100): 12.8 events/3.5\,d = 3.7/d
\end{enumerate}

The total number of events for the above series is $(26.8 \pm 5.4)$ events in 5.9\,d, or $(4.5 \pm 0.9)$/d. The uncertainty includes the small background subtraction. Full production running (reactor cycles 3 through 8) for the BGO detectors operated for 85.8 days and registered about 44600 events. Comparing these test runs and the production runs, we expect about 1\,\% of false correlated events could be included in our data set.  We therefore correct the branching ratio by $-1$\,\% and assign 100\,\% of the correction as the systematic uncertainty. The comparable analysis for the APD detectors gives a correction of $-0.4$\,\%.

\subsection{Simulation}

\subsubsection{Model registration}
\label{sys:model}

Because the branching ratio is formed as a ratio of the experimental to the MC simulation, the fidelity of the input to the MC is critical. The relative positions of detectors, magnetic field, and beam with respect to each other were of particular importance. Care was taken to physically measure and map their positions in the apparatus. The registration uncertainties were converted to uncertainty in the branching ratios by varying the physical positions in the simulation. Table~\ref{tab:model} lists the largest uncertainty contributions due to registration shifts. The final uncertainty due to model registration was made by adding the individual contributions in quadrature.

For this analysis, the coordinate system is such that {\it z} is the direction along the beam axis and {\it x} (horizontal) and {\it y} (vertical) are in the plane perpendicular to the beam axis.
Shifts in the {\it x} direction resulted in large differences in $ep$ detection, which in turn affect the $ep\gamma/ep$ efficiency and the branching ratio.  These were found in the simulation to originate from decays occurring in the bend region of the magnet (see Fig.~\ref{fig:diagram}) where some of the electrons or protons could strike insensitive regions of the SBD or completely miss the detector. Shifts along the {\it z}-axis of the beam  could also produce significant effects in the branching ratio. The effect in the {\it y}-axis was much less due to the symmetry in that orientation, and it contributed negligibly.

\begin{table}[h]
\caption{\label{tab:model} Uncertainties due to the registration of components in the MC model of the apparatus. ``Detector'' refers to the assembly with the BGO and APD detectors and the mirror.}
\begin{ruledtabular}
\begin{tabular}{l|ccc}
Component           & Shift (mm)    & BGO Unc. (\%) & APD Unc. (\%)   \\
\hline
Beam ({\it x})      & $\pm 1$       & 1.9       & 1.8       \\
Beam ({\it z})      & $\pm 1$       & 0.2       & 0.7       \\
B field ({\it x})   & $\pm 1$       & 0.3       & 0.8       \\
B field ({\it z})   & $\pm 1$       & 0.2       & 0.7       \\
SBD ({\it x})       & $\pm 1$       & 1.9       & 1.8       \\
SBD ({\it z})       & $\pm 2$       & 0.3       & 1.5       \\
Detector ({\it x})  & $\pm 1$       & 0.2       & 0.8       \\
Detector ({\it z})  & $\pm 1$       & 0.3       & 0.8       \\
\hline
Total               &               & 2.8       & 3.4       \\
\end{tabular}
\end{ruledtabular}
\end{table} 

The registration shifts were 1\,mm, which corresponds to the best that the components and fields of the apparatus could be measured confidently. The exception was the z direction of the detector array; a shift of $\pm 2$\,mm was used to account for uncertainty in the position from thermal contraction as the detector cooled from room temperature to tens of kelvin. 

\subsubsection{Charged particle backscattering}
\label{sys:backscatter}

A small fraction of the protons and electrons striking the surface of the SBD will backscatter. For the $-25$\,keV incident protons that backscatter, some may return to the detector and be registered (but with a reduced energy), and others may be completely lost. In either case, proton backscattering would affect the overall efficiency of proton detection but have a negligible effect on the branching ratio or the photon energy spectrum.

Electron backscattering was included in the GEANT simulation. It affects the electron energy spectrum but has little effect on the branching ratio and the photon energy spectrum. Several low-energy libraries of \textsc{Geant4.9.6.p02} were compared, and it was found that, even when single scattering was used, no significant differences were found in the number of counts.


\section{Conclusion}
\label{sec:conclusion}

\subsection {Results}
\label{subsec:results}

Decisions on the analysis methods were made to make the most meaningful comparison between the experimental data and simulation. For the $ep\gamma$ coincidences, it was necessary to determine how to combine data from the individual detectors for the BGO and APD arrays. Two detector combination methods were considered: one, adding the peak heights from each detector in the array for each $ep$ trigger (treating the array as a single photon detector), and two, analyzing each detector individually and averaging their spectra. The latter method was selected to analyze the data because it does not require uniformity of the entire detector array. For example, if a detector were not functioning properly during a series, it would produce complications in the photon response arising from the bismuth x-rays from inactive BGO crystals affecting the energy spectrum near 80\,keV in operational crystals. This problem is greatly mitigated by treating each detector as its own experiment.

There are also two techniques of extracting a radiative branching ratio. One technique
is a $\chi^2$ fit to the energy spectrum by allowing the branching ratio to float as a scaling constant. Another approach compares only the integral number of counts within a photon energy range when determining a branching ratio. Analyses were done using both techniques and no significant difference was found. For systematic reasons the latter method was preferred as counting photons above a threshold is less susceptible to issues associated with the spectral shape.

\begin{figure*}
 \includegraphics[width=0.49\textwidth]{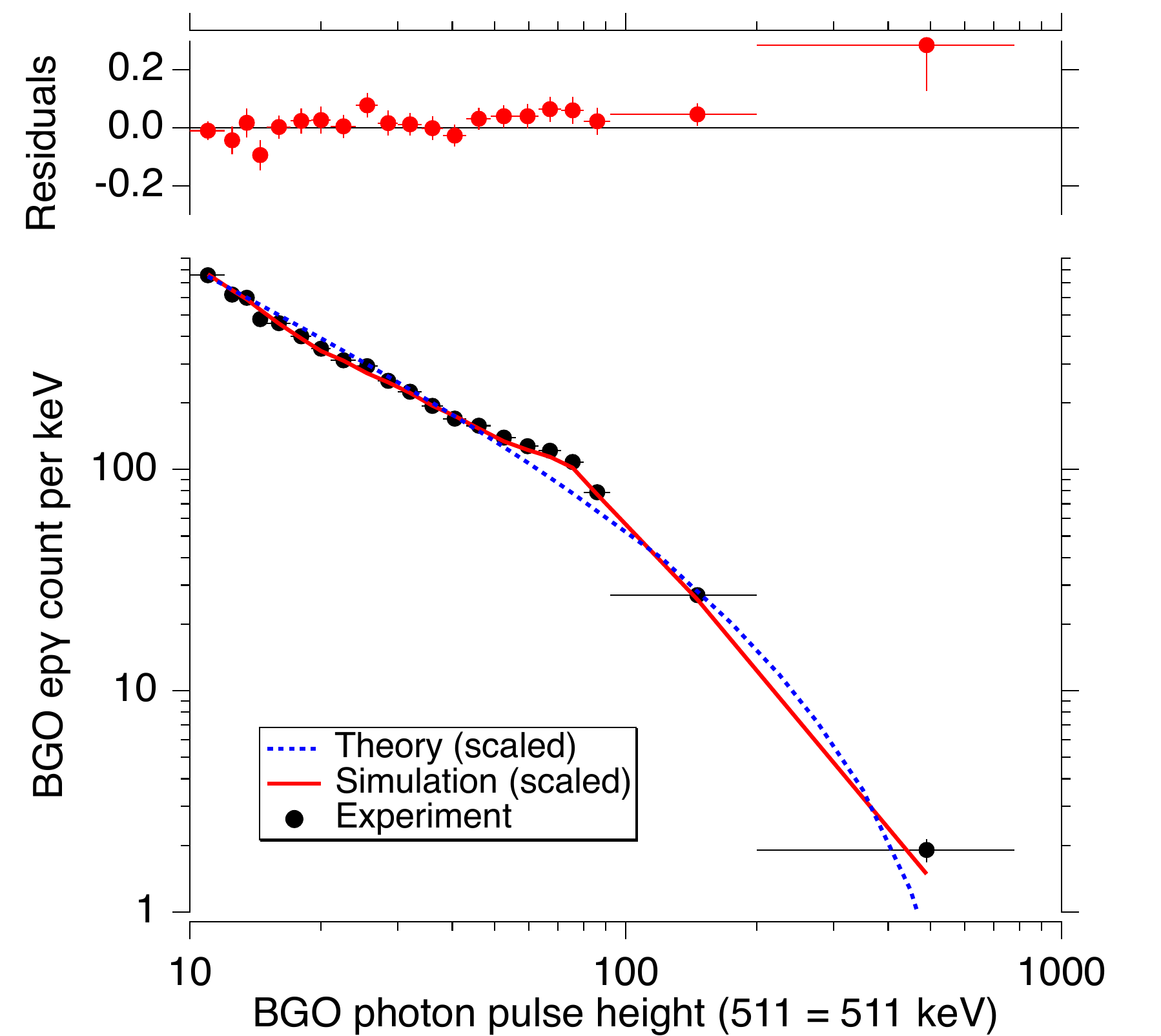}
 \includegraphics[width=0.49\textwidth]{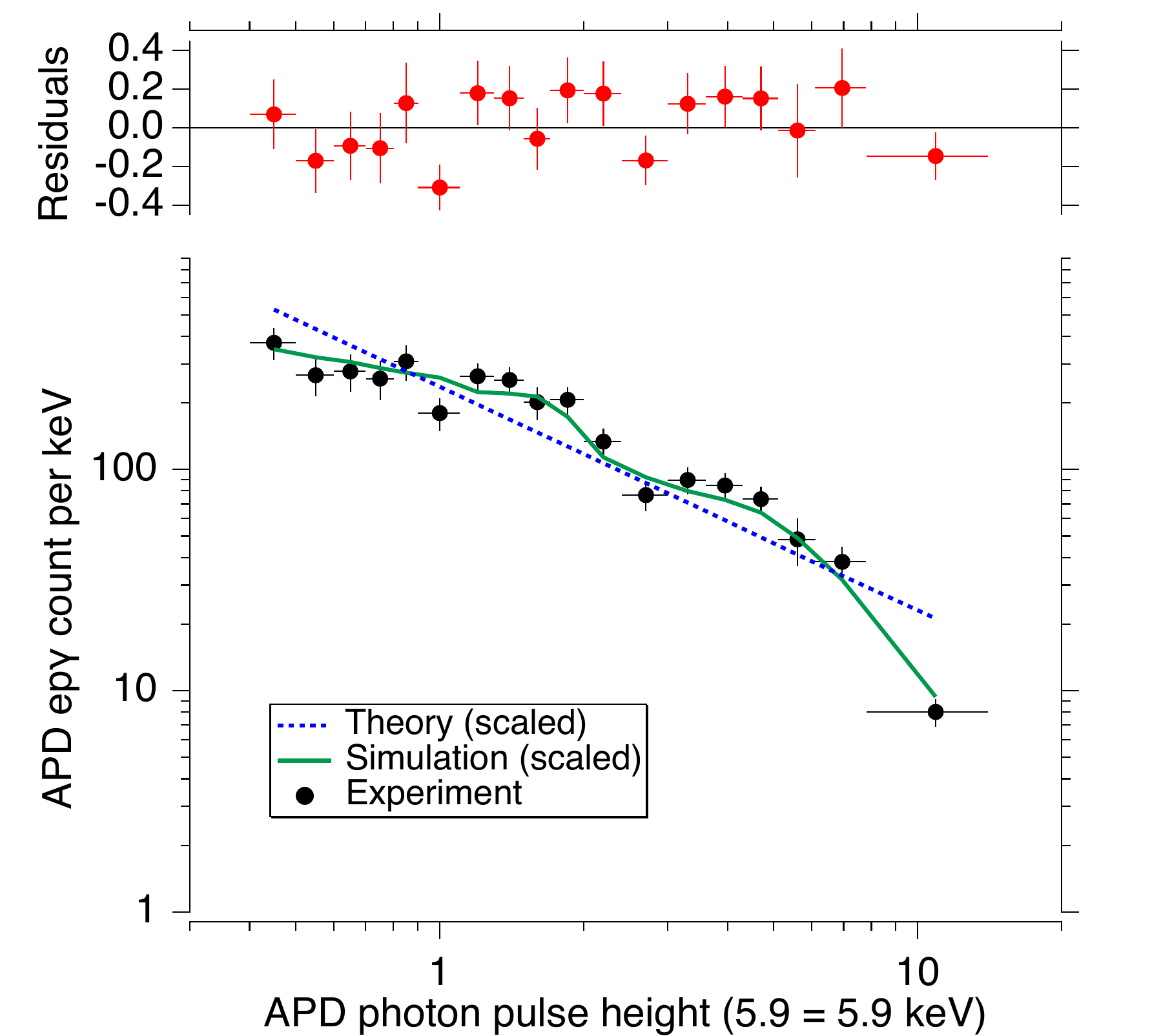}
 \caption
{\label{fig:spectra}
Energy spectrum deposited by photons from radiative neutron decay. The average background-corrected radiative photon counts for both the BGO (left) and APD (right) detectors versus photon pulse height are plotted. The bump at 80\,keV in the BGO spectrum arises from a bismuth x-ray and is discussed in the text. The experimental data (black circles) include only the statistical uncertainty in the vertical error bars while the horizontal error bars represent the bin size. The blue dashed lines show the theoretical spectrum scaled to the experimental data and plotted versus photon energy. The solid lines are the output of the simulation scaled to the experimental data using the theoretical spectrum as input; the simulation results are binned in energy to match the data.   The normalized residuals (black circles) between the experiment and simulation are shown at the top.  The vertical error bars represent the statistical uncertainties only.}
\end{figure*}

The cuts from Table~\ref{tab:cuts} were applied to the data in the series of the final data set, and the appropriately normalized background windows from Fig.~\ref{fig:egTime} were subtracted from the photon energy spectra. This was done for each individual detector for both the BGO and APD detectors. For the final result, the radiative spectrum from each individual detector was averaged, and the branching ratio was determined from Eq.~(\ref{eqn:result_eqn}). Fig.~\ref{fig:spectra} shows the combined energy spectra for the BGO and APD arrays. The simulation incorporates the coincident detection of the decay particles and the response functions for the arrays. The x-axes are the photon energies based on the calibrations discussed in Section~\ref{sys:photoncal}.  Note that photons in the low energy region are registered below the $x$-axis energy for each detector, either due to nonproportionality of the BGO crystals or decreased collection efficiency of electron-hole pairs in the APDs.

For the BGO array, the averaged energy spectrum agrees well with the scaled average spectrum predicted by simulation and results in a chi-squared per degree of freedom of 20.2/20 with a p-value of 0.44. The bump at $\approx$80 keV is caused by the escape of bismuth K-shell X-rays from nearby crystals.  The branching ratio $B_{\textrm{Exp}}^{\textrm{BGO}}$ was measured to be 0.00335 $\pm$ 0.00005 [statistical] $\pm$ 0.00015 [systematic] in the energy range of 14.1\,keV to 782\,keV. In this range, $B_{\textrm{Theory}}^{\textrm{BGO}}$ was calculated to be 0.00308. The values  $B_{\textrm{Exp}}^{\textrm{BGO}}$ and $B_{\textrm{Theory}}^{\textrm{BGO}}$ agree within $1.7$ times the combined standard uncertainty.

The APD data were analyzed similarly. The averaged energy spectrum agrees well with the scaled average spectrum predicted by simulation and yielded a chi-squared per degree of freedom of 20.1/17 with a p-value of 0.27. The shoulder just below 2\,keV is due to the order-of-magnitude decrease in the attenuation length that occurs for energies above the 1.8\,keV K-edge in silicon~\cite{Gentile2012}. The branching ratio $B_{\textrm{Exp}}^{\textrm{APD}}$was measured to be 0.00582 $\pm$ 0.00023 [stat] $\pm$ 0.00062 [syst] in the range of 0.4\,keV to 14\,keV. In this range, $B_{\textrm{Theory}}^{\textrm{APD}}$ was calculated to be 0.00515. The values $B_{\textrm{Exp}}^{\textrm{APD}}$ and $B_{\textrm{Theory}}^{\textrm{APD}}$ agree within $1.0$ times the combined standard uncertainty.

The branching ratios in both energy ranges are in agreement with QED predictions, and the energy spectra fit theory well over three decades of energy. The corrections and relative standard uncertainties of systematic effects associated with both the BGO and APD measurements are summarized in Table~\ref{table:Systematics}. The results do not differ significantly from the original publication~\cite{Bales2016}.  The dominant systematic uncertainties in this experiment were in the simulation's model registration, pulse shape discrimination, and photon detector energy response. Corrections were made for bremsstrahlung induced by electrons interacting with the SBD and correlated but non-decay related backgrounds. 

\subsection{Future Work}
\label{subsec:future}

As discussed in Section~\ref{sec:overview}, there are several motivations for improving the precision in measurements of radiative neutron decay. With the experience gained in this experiment, it is possible to design an apparatus that can achieve significantly reduced statistical and systematic uncertainties. The approach one would take and how the detectors are designed depend greatly on the physics being probed. One needs to consider whether the goal is better precision in the branching ratio, improved measurement of the energy spectrum, or measuring angular correlations or the photon's circular polarization. Regardless, the precision is currently limited by systematic effects, so a significantly better understanding of these effects is essential.

There are two general approaches that could be taken. This experiment used a large magnetic field to increase the collection efficiency of the charged decay products. We collected about half of the decay electrons and, with the electrostatic mirror, nearly all of the protons. The disadvantage of this approach is that the experiment sacrifices the angular information of the decay products, and thus it is impossible to generate correlations among the decay products. However, the method improves the counting statistics and has no detrimental effect on the measured energy spectrum.

One could imagine a straightforward scaling of this experiment. The fabrication of such an apparatus is already in progress by the BL3 collaboration~\cite{BL32021} to measure the neutron lifetime. The design of the magnet and proton detection system is similar to that used in this experiment, but it is larger in scale and incorporates some technological improvements. The bore of the magnet is approximately 50\,\% larger in diameter and length, allowing the use of a larger diameter beam and a larger detector array. With a scaled-up detector and a beam diameter four times larger, one could achieve about a factor of 4 improvement in statistics for the same live time as for RDK II.

It is still necessary to address the limiting systematic effects, the largest of which is the energy response of the photon detector and registration of the relevant components of the apparatus. For the energy response, other inorganic scintillators  with better linearity should be investigated for relevant properties such as vacuum compatibility, cryogenic operation, and light yield and photopeak efficiency at low temperature. The systematic for the physical registration of the apparatus should be greatly improved by using the existing metrology of the BL3 magnet and designing the detector components to fit kinematically. With these improvements, it should be possible to improve the overall precision in the branching ratio to less than 1\,\%.

A second approach can be considered that eliminates the large magnetic field, thus allowing particle tracking and an improved detection-volume definition. There are already experiments that use the reconstruction of the neutron decay products to measure decay correlation coefficients~\cite{Lising2000,Kozela2009}. Quantum sensors are currently being developed to measure charged particles that may yield significantly better energy resolution, and hence better particle identification. The technology is not yet mature enough that an experiment could be designed around quantum sensors, but the field is moving quickly and should be considered for future beta decay experiments. 

\begin{acknowledgments}
We thank Changbo Fu for his initial simulation work and David Winogradoff for his calibration studies.  We additionally thank R. Farrell for numerous discussions about APD operation and their properties. This research was supported in part through computational
resources provided by The University of Michigan, Ann Arbor, Advanced Research
Computing services. We acknowledge the support of the National Institute of Standards and Technology, US Department of Commerce, in providing the neutron facilities used in this work. This research was made possible in part by support from the National Science Foundation (PHY-1205266, PHY-1205393, PHY-1306547, PHY-1505196, and PHY-2309938) and the US Department of Energy (DE-FG02-96ER40989 and an interagency agreement).

Identification of a product herein is for documentation purposes only, and does not imply recommendation or endorsement by NIST, nor does it imply that this product is
necessarily the best available for the purpose.

\end{acknowledgments}
\bibliography{RDKIIPRC}

\end{document}